\documentclass[conference,10pt]{IEEEtran}

\usepackage{framed}
\usepackage{algorithm}
\usepackage[noend]{algpseudocode}
\usepackage{bm}
\usepackage{stackrel}
\usepackage{subfigure}
\usepackage{epsf}
\usepackage{amsmath,amssymb}
\usepackage{graphicx}
\usepackage{color}
\usepackage{cite}
\usepackage{multirow,tabularx}
\usepackage{ifthen}
\usepackage{epstopdf}
\input{epstopdf.sty}
\usepackage{times}

\input{epsf.sty}
\makeatletter
\def\BState{\State\hskip-\ALG@thistlm}
\makeatother

\newcommand{\hide}[1]{\ifthenelse{\boolean{false}}{#1}{}}

\newtheorem{theorem}{{\bf Theorem}}

\newtheorem{lemma}{{\bf Lemma}}

\newcommand{\qed}{\nobreak \ifvmode \relax \else
      \ifdim\lastskip<1.5em \hskip-\lastskip
      \hskip1.5em plus0em minus0.5em \fi \nobreak
      \vrule height0.75em width0.5em depth0.25em\fi}

\newcommand{\beq}{\begin{equation}}
\newcommand{\eeq}{\end{equation}}
\newcommand{\barr}{\begin{array}}
\newcommand{\earr}{\end{array}}

\newcommand{\benum}{\begin{enumerate}}
\newcommand{\eenum}{\end{enumerate}}

\newcommand{\bit}{\begin{itemize}}
\newcommand{\eit}{\end{itemize}}

\newcommand{\bc}{\begin{center}}
\newcommand{\ec}{\end{center}}

\newcommand{\bdes}{\begin{description}}
\newcommand{\edes}{\end{description}}

\newcommand{\bfig}{\begin{figure}}
\newcommand{\efig}{\end{figure}}

\newcommand{\bemq}{\begin{quote} \begin{em}}
\newcommand{\eemq}{\end{em} \end{quote}}

\newcommand{\bmp}{\begin{minipage}}
\newcommand{\emp}{\end{minipage}}

\newcommand{\EX}[1]{\mathbb{E}\left[{#1}\right]} % expectation operator

\newcommand{\bsp}{\begin{slide*}}
\newcommand{\esp}{\end{slide*}}
\newcommand{\bsl}{\begin{slide}}
\newcommand{\esl}{\end{slide}}

\newcommand{\blem}{\begin{lemma}}
\newcommand{\elem}{\end{lemma}}
\newcommand{\bthm}{\begin{theorem}}
\newcommand{\ethm}{\end{theorem}}

\newcommand{\AoI}{\text{AoI}}
\newcommand{\VarD}{\text{VarD}}

\newcommand{\pr}[1]{\mathbf{P}\left[ #1 \right]}

\IEEEoverridecommandlockouts

\begin{document}

%\title{Age-Delay Tradeoffs in $M$ Server Systems}
\title{Age-Delay Tradeoffs in Queueing Systems}
%\title{Optimizing Age of Information over Update Generation and Service Time Distributions}
%\title{Can Determinacy Reduce Age of Information?}
%\title{Effect oDeterminacy in Arrival and Service of Updates on Age of Information}
% \date{\today}
\author{Rajat Talak and Eytan Modiano
\thanks{This work was supported by NSF Grants AST-1547331, CNS-1713725, and
CNS-1701964, and by Army Research Office (ARO) grant number W911NF-
17-1-0508.}
\thanks{A preliminary version of the this work was available on arXiv~\cite{talak18_determinacy} and appeared in ISIT 2019~\cite{talak19_AoI_age_delay, talak19_AoI_heavytail}.}
\thanks{The authors are with the Laboratory for Information and Decision Systems (LIDS) at the Massachusetts Institute of Technology (MIT), Cambridge, MA. {\tt \{talak, modiano\}@mit.edu} }
}

\IEEEaftertitletext{\vspace{-0.6\baselineskip}}

\maketitle

\begin{abstract}
We consider an M server system in which each server can service at most one update packet at a time. The system designer controls (1) \emph{scheduling} - the order in which the packets get serviced, (2) \emph{routing} - the server that an arriving update packet joins for service, and (3)~the \emph{service time distribution} with fixed service rate. Given a fixed update generation process, we prove a strong age-delay and age-delay variance tradeoff, wherein, as the average AoI approaches its minimum, the packet delay and its variance approach infinity. In order to prove this result, we consider two special cases of the M server system, namely, a single server system with last come first server with preemptive service and an infinite server system. In both these cases, we derive sufficient conditions to show that three heavy tailed service time distributions, namely Pareto, log-normal, and Weibull, asymptotically minimize the average AoI as their tail gets heavier, and establish the age-delay tradeoff results. We provide an intuitive explanation as to why such a seemingly counter intuitive age-delay tradeoff is natural, and that it should exist in many systems.

\end{abstract}

\section{Introduction}
\label{sec:intro}
Information freshness and low latency communication is gaining increasing relevance in many futuristic communication systems, such as industrial automation, autonomous driving, tele-surgery, financial markets, and virtual reality~\cite{2018_LowLatencySurvey_Fischione, 2016_5G_Enabled_Tactile_Internet, 2018_Haptic_and_5G, 2016_Tactile_Internet_Vision_Progress_Challenges}. The latency requirements vary depending on the application. While applications such as autonomous driving, tele-surgery, virtual reality, financial markets are envisioned to require a latency of a few milliseconds, other systems such as industry automation, control signalling in power grids aim at a latency of 10-100 milliseconds~\cite{2018_LowLatencySurvey_Fischione, 2016_5G_Enabled_Tactile_Internet}. In many of these applications, seeking the most recent status update is crucial to the overall system performance. In a network of unmanned aerial vehicles, for example, exchanging the most recent position, speed, and other control information~\cite{talakCDC16, FANETs2013}; in operations monitoring systems, accessing the most recent sensor measurement; and in cellular systems, obtaining the timely channel state information from the mobile users~\cite{LTE_book}, is important and can lead to significant performance improvements.

\begin{figure}
  \centering
  \includegraphics[width=0.95\linewidth]{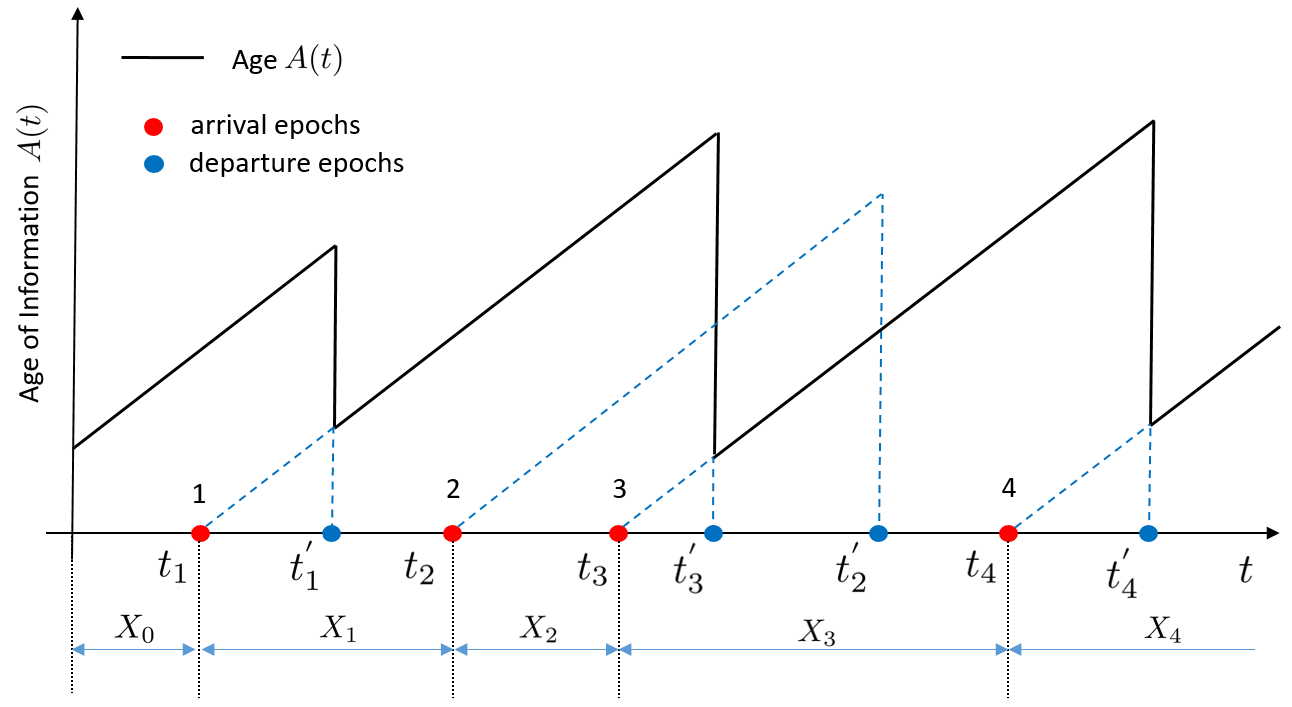}
  \caption{Age evolution in time. Update packets generated at times $t_i$ and received, by the destination, at times $t^{'}_{i}$. Packet $3$ is received out of order, and thus, doesn't contribute to age.}
  \label{fig:age}
\end{figure}
Age of information (AoI) is a metric for information freshness that measures the time that elapses since the last received fresh update was generated at the source~\cite{2011SeCON_Kaul, 2012Infocom_KaulYates}. It is, therefore, a destination-centric measure, and is suitable for applications that necessitate timely updates. A typical evolution of AoI for a single source-destination system is shown in Figure~\ref{fig:age}. The AoI increases linearly in time, until the destination receives a fresh update packet. Upon reception of a fresh update packet $i$, at time $t^{'}_{i}$, the AoI drops to the time since the packet $i$ was generated, which is $t^{'}_i - t_i$; here $t_i$ is the time of generation of the update packet $i$. Unlike the traditional latency metrics such as packet delay, AoI only accounts for the update packets that deliver fresh updates to the destination -- such packets are called \emph{informative packets}~\cite{2014ISIT_KamKomEp}. For example, in Figure~\ref{fig:age}, packet $3$ is an informative packet but packet $2$ is not. This is because packet $3$ reaches the destination before packet $2$, which is therefore rendered stale by time $t^{'}_{2}$.

AoI was first studied for the first come first serve (FCFS) M/M/1, M/D/1, and D/M/1 queues in~\cite{2012Infocom_KaulYates}. Since then, AoI has been analyzed for several queueing systems with the goal to minimize AoI~\cite{2011SeCON_Kaul, 2012CISS_KaulYates, 2012Infocom_KaulYates, 2012Infocom_TIT_KaulYates, 2015ISIT_LongBoEM, talak17_allerton, talak18_Mobihoc, talak19_ToN_Mobihoc, vishrant19_AoI_discreteQ, 2016X_Najm, 2016_ISIT_YinSun_AoI_Thput_Delay_LCFS, 2016_ISIT_TIT_YinSun_Thput_Delay_LCFS, 2017_ISIT_YinSun_LCFSopt_MultiHop, 2017_ISIT_TIT_YinSun_LCFSopt_MultiHop, 2014ISIT_KamKomEp, 2014ISIT_CostaEp, Inoue17_FCFS_AoIDist, 2018_Ulukus_GG11, 2018ISIT_Yates_AoI_ParallelLCFS, 2018_Yates_LCFS_Multihop, 2016_ISIT_Ep_AoI_Deadlines, 2018_ISIT_Inoue_AoI_Deadline, 2016_MILCOM_Ep_AoI_Buffer_Deadline_Replace, YinSun_2019_AoI_book}. Two time average metrics of AoI, namely, peak and average age are generally considered. The analysis mostly relies on the specificities of the queueing model under consideration. Typically, a peak or average age expression is derived and then optimized over the update generation rate. However, progress has been made recently towards a more general analysis of AoI. A general formula for the stationary distribution of AoI for a single-server systems was recently developed in~\cite{Inoue17_FCFS_AoIDist, Inoue18_FCFS_LCFS_AoIDist}, while~\cite{2018_Yates_SHS} used the theory of stochastic hybrid systems to systematically derive expressions for the average age and its higher moments.

%AoI for the LCFS queue with Poisson arrivals and Gamma distributed service was analyzed in~\cite{2016X_Najm}. Average age for a series of LCFSp queues in tandem was analysed in~\cite{talak17_allerton, 2018_Yates_LCFS_Multihop}. Complexity of extending the traditional queuing theory analysis to analyzing multi-hop, multi-server systems has lead~\cite{2018_Yates_SHS} to propose stochastic hybrid system method to compute average age, and its moments.
%
%The analysis mostly relies on the specificities of the queueing model under consideration.
%
% Peak age for FCFS G/G/1, M/G/1 and multi-class M/G/1 queueing systems was analyzed in~\cite{2015ISIT_LongBoEM}, while the discrete time queues were studied in~\cite{talak18_Mobihoc, talak19_ToN_Mobihoc, vishrant19_AoI_discreteQ}. A general formula for the stationary distribution of AoI for single-server systems has been recently developed in~\cite{Inoue17_FCFS_AoIDist, Inoue18_FCFS_LCFS_AoIDist}. %Age for preemptive and non-preemptive last come first serve (LCFS) queue with Poisson arrival and Gamma distributed service was analyzed in~\cite{2016X_Najm}.

Despite the difficulty in analyzing age for general queueing systems several approaches that reduce or minimize AoI have been brought to light. The advantage of having parallel servers in reducing AoI was demonstrated in~\cite{2014ISIT_KamKomEp, 2014ISIT_CostaEp, 2018ISIT_Yates_AoI_ParallelLCFS}. Methods such as limiting the buffer sizes~\cite{2011SeCON_Kaul, 2016_MILCOM_Ep_AoI_Buffer_Deadline_Replace} or introducing packet deadlines~\cite{2016_MILCOM_Ep_AoI_Buffer_Deadline_Replace, 2016_ISIT_Ep_AoI_Deadlines, 2018_ISIT_Inoue_AoI_Deadline} have also been shown to reduce AoI. In a general queueing system, with exponentially distributed service times, the last come first serve (LCFS) queue scheduling discipline with preemptive service was proved to be age optimal in~\cite{2016_ISIT_YinSun_AoI_Thput_Delay_LCFS, 2016_ISIT_TIT_YinSun_Thput_Delay_LCFS, 2017_ISIT_YinSun_LCFSopt_MultiHop, 2017_ISIT_TIT_YinSun_LCFSopt_MultiHop}.  %\rt{Add other literatures on age in one-to-two lines.}
In~\cite{yin17_tit_update_or_wait}, an optimal update generation policy was investigated, and it was discovered that an intuitively apt zero-wait policy, which sends the next update right after the previous one is received, is not always age optimal. %and it is found that sometimes it is optimal to wait a while, even after the current update has been received by the destination, before sending the next update.

More recently, minimizing age metrics over update generation and service time distribution has been of interest~\cite{talak18_determinacy, Inoue18_FCFS_LCFS_AoIDist}. In a related work~\cite{talak18_determinacy}, we considered the problem of minimizing peak and average age over packet generation and service time distributions, given a particular update generation and service rate. We showed that determinacy in packet generation and/or service does not necessarily minimize age. Similar results were independently obtained in~\cite{Inoue18_FCFS_LCFS_AoIDist}.

%In all these status update systems, the generation of update packets is generally under the control of the system designer. As a consequence, in most of these works, the peak or average age expression is obtained, and an optimal packet generation rate is sought, that minimizes the respective age metric. Service time distribution generally depends on the packet length, and therefore, a given packet length distribution can be induced on the generated update packets. In a related work, we optimize peak and average age over packet generation and service time distributions, given a particular update generation and service rate~\cite{talak18_determinacy}. We show that determinacy in packet generation and/or service does not necessarily minimize age. Similar results were independently obtained in~\cite{Inoue18_FCFS_LCFS_AoIDist}.

Packet delay and delay variance (jitter), on the other hand, have traditionally been considered as measures of communication latency. Minimizing packet delay in a queueing system is known to be a hard problem. Several works have focused on the problem of reducing or minimizing the packet delay and its variance~\cite{1962_Kingman_Q_Discipline_Wait_Variance, 1975_Rolski_Stoyan_Comparison_Waiting_Times_GG1, 1977_Wolff, 1980_Whitt, 1984_Whitt, 1983_Hajek, 1983_Whitt, 2019_Ness_DelayOptimal_EnergyEffi_Commun_MarkovArrivals, 2017_PingChun_IHong_DelayOptimalityQueues_Switching_Overhead, 2017_Ness_FastConvergent_LowDelay_LowComplexity_NetwOpt, 2016_YinSun_Ness_DelayOptimal_Scheduling_Queues_Pkt_Replications, 2016_Infocom_HeavyBall_to_Tame_Delay, 2013_Neely_Delay_Based_NUM, 2013_Ness_Delay_Based_MaxWeight, 2013_Neely_LIFO_Backpressure, 2011_Ness_DelayAnalysis_MultiHop}. It is widely believed that AoI and delay are closely related, and hence, can be minimized simultaneously. For example, in a simple FCFS queue under Poisson arrivals, less variability in service time distribution minimizes both packet delay and peak age~\cite{talak18_determinacy, Inoue18_FCFS_LCFS_AoIDist}. For a system of $M$ parallel servers with exponential service times, minimum age and delay can be simultaneously attained by resorting to the LCFS with preemptive service~\cite{2016_ISIT_YinSun_AoI_Thput_Delay_LCFS, 2016_ISIT_TIT_YinSun_Thput_Delay_LCFS}.

\emph{Is it then always possible to minimize age and packet delay simultaneously, or are there systems in which minimizing one does not imply minimizing the other?}

In this work, we answer this question by considering an $M$ server queueing system. We show that as we tailor the queue scheduling discipline, routing, and the service time distributions to minimize average age, the packet delay and its variance approach infinity.
\begin{figure}
  \centering
  \includegraphics[width=0.95\linewidth]{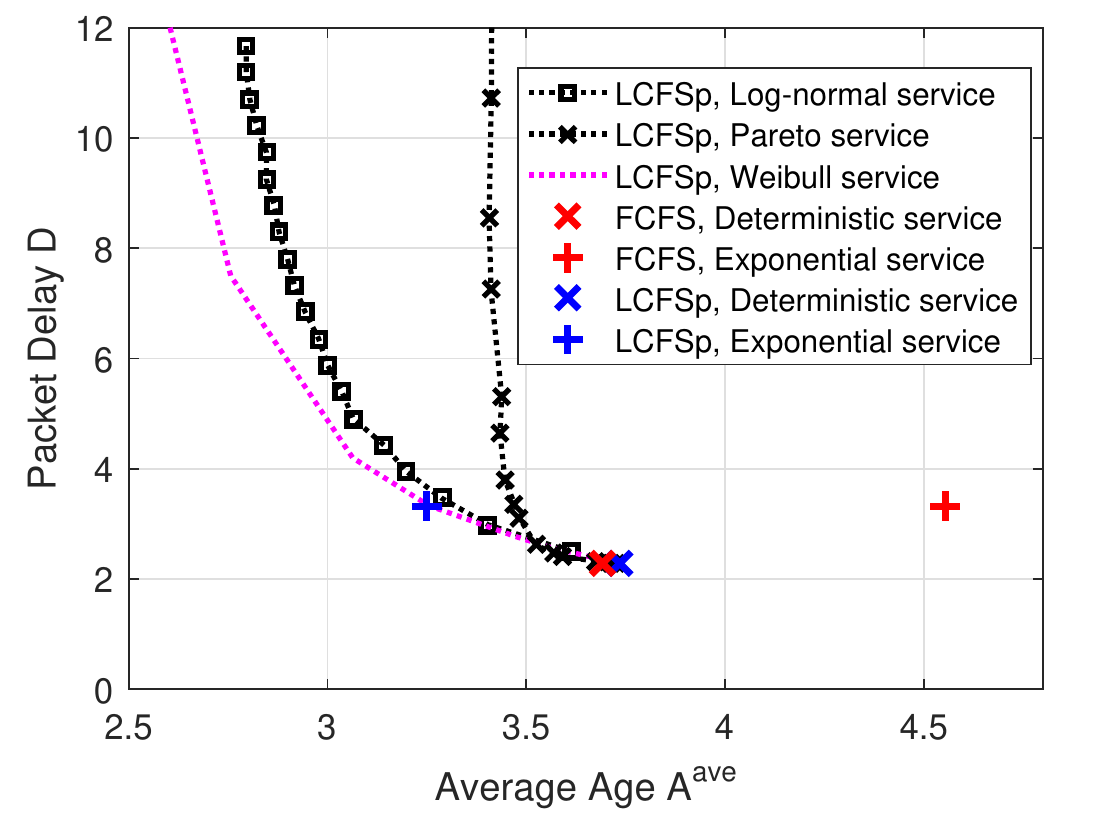}
  \caption{Plot of achieved age-delay points for various single server systems, Poisson packet generation at rate $\lambda = 0.5$, and service at rate $\mu = 0.8$. Scheduling disciplines: FCFS, LCFS with preemptive service. Service time distributions: Deterministic, Exponential, and Heavy Tailed distributions in Table~\ref{tbl:heavy_tail}.}
  %\caption{Plot of packet delay as a function of average age for single server systems, under various scheduling disciplines (FCFS and LCFSp) and service time distributions (Deterministic, Exponential, and Heavy Tailed distributions in Table~\ref{tbl:heavy_tail}). Packet generation is Poisson at rate $\lambda = 0.5$ and service rate $\mu = 0.8$.}
  \label{fig:AoI_Delay_tradeoff1_plot1}
\end{figure}
As an example, consider a single server queue with a fixed update generation and service rate. Updates are generated according to a Poisson process. In Figure~\ref{fig:AoI_Delay_tradeoff1_plot1}, we plot the achieved packet delay and average age attained under various queue scheduling policies (FCFS and LCFS with preemptive service) and service time distributions (deterministic, exponential, log-normal, Pareto, and Weibull). For the three heavy tailed service time distributions, namely, log-normal, Pareto, and Weibull we plot the age-delay points for various values of the free parameter; see Table~\ref{tbl:heavy_tail}. It appears that there is a strong age-delay tradeoff, i.e. a lower average age can be achieved at the cost of higher delay.

Intuitively, this tradeoff can be explained as follows: \emph{In order to achieve minimum age the system has to prioritize informative packets. In doing so, the non-informative updates lag behind in the system thereby incurring a large waiting time cost. The delay and its variance get dominated by the large delays incurred by the non-informative update packets, thus leading to the tradeoff curve in Figure~\ref{fig:AoI_Delay_tradeoff1_plot1}.}

Given this intuition, we suspect that an age-delay tradeoff should exist in many systems. In this paper, we prove it for an $M$ server queueing system.

\begin{table}
\caption{Heavy tailed service time distributions with mean $\EX{S} = 1/\mu$.}
\label{tbl:heavy_tail}
\begin{center}
\begin{tabular}{ |c|c|c| }
 \hline
 Name & Distribution & Free Parameter \\
 \hline
 Log-normal & $S = \exp\left(- \log \mu - \frac{\sigma^2}{2} + \sigma N\right)$ & $\sigma > 0$ \\
 & $N \sim \mathcal{N}(0, 1)$ & \\
 \hline
 Pareto & $F_S(x) = 1 - \left(\theta(\alpha)/x\right)^{\alpha}\mathbb{I}_{\{ x > \theta(\alpha) \}}$  & $\alpha > 1$ \\
 & $\theta(\alpha) = (\alpha - 1)/(\mu \alpha)$ & \\
 \hline
 Weibull & $\pr{S > x} = \exp\left\{- (x/\beta(k))^{k} \right\}$ & $k > 0$ \\
 & $\beta(k) = \left[ \mu \Gamma(1 + 1/k)\right]^{-1}$ & \\
 \hline
\end{tabular}
\end{center}
\end{table}

\subsection{Contributions}
We consider an $M$ server system in which each server can service at most one update packet at any given time. Update packets are generated according to a renewal process at a fixed rate. The system designer decides the queue \emph{scheduling discipline}, i.e. the order in which the packets get serviced, the \emph{routing}, which determines the server for each arriving update packet, and the \emph{service time distribution}. %Note that the service time distribution generally depends on the packet length, and therefore, a given packet length distribution can be induced on the generated update packets.

In order to observe the age-delay tradeoff, we consider the problem of minimizing packet delay (and packet delay variance), subject to an average age constraint, over the space of all queue scheduling disciplines, routing, and service time distributions, with a fixed mean service time budget of $1/\mu$ for each queue. When the updates are generated according to a Poisson process we show that there is a strong age-delay tradeoff, namely, as the average age approaches its minimum, the delay approaches infinity. When the updates are generated according to a general renewal process, we show that there is a strong age-delay variance tradeoff, i.e. as the average age approaches its minimum, the variance in packet delay approaches infinity.

The proof of this result involves first proving the same result in two special cases of the $M$ server system: (1)~A single server system, i.e. $M = 1$, in which the queue scheduling discipline is fixed to LCFSp, and (2)~An infinite server system, i.e. $M = \infty$. In both these cases, we derive sufficient conditions on the average age minimizing service time distribution. We then show that these sufficient conditions are satisfied by the three heavy tailed service time distributions, namely Pareto, log-normal, and Weibull, asymptotically in its tail parameter. This helps us establish the age-delay tradeoff results in the two special cases of the $M$ server system. We also observe a certain age-delay disparity in these two cases in which the delay (or delay variance) minimizing service time distributions result in the worst case average age.

The results derived for the two special cases are then used to prove the strong age-delay and age-delay variance tradeoffs for the $M$ server system. To the best of our knowledge, this is the first work to establish an age-delay tradeoff result. A preliminary version of the this work was available on arXiv~\cite{talak18_determinacy} and appeared in ISIT 2019~\cite{talak19_AoI_age_delay, talak19_AoI_heavytail}. This work builds upon the results in~\cite{talak19_AoI_age_delay, talak19_AoI_heavytail, talak18_determinacy}.

%\subsection{Related Works}
%\label{sec:lit}

\subsection{Organization}
\label{sec:org}
In Section~\ref{sec:model}, we describe the system model and provide a general definition of AoI. In Section~\ref{sec:age_delay_tradeoff}, we formulate the age-delay and age-delay variance problems for the $M$ server system. The age-delay tradeoff result for the $M$ server system is also stated and discussed here. In Sections~\ref{sec:lcfs} and~\ref{sec:inf_serv}, we prove the age-delay tradeoff result in the two special cases of the single server LCFSp and infinite server systems.
The paper culminates in Section~\ref{sec:M_ServSyst} with a proof of the age-delay tradeoff for the $M$ server system. We conclude in Section~\ref{sec:conclusion}.
%In Section~\ref{sec:M_ServSyst}, the paper culminates in a proof of the age-delay tradeoff for the $M$ server system. We conclude in Section~\ref{sec:conclusion}.

%The system model and a generic definition of AoI is given in Section~\ref{sec:model}. %and provide a generic definition of AoI and average age.
%
%Age-delay and age-delay variance tradeoff problems are set up, and the age-delay tradeoff for the $M$ server system - the central result in this paper - is stated in Section~\ref{sec:age_delay_tradeoff}. Sections~\ref{sec:lcfs} and~\ref{sec:inf_serv} establishes the age-delay tradeoff result in the two special cases of the single server LCFSp and infinite server systems. Finally, in Section~\ref{sec:M_ServSyst}, we prove the age-delay tradeoff for the $M$ server system. We conclude in

%\pagebreak
\section{System Model}
\label{sec:model}

\begin{figure}
  \centering
  \includegraphics[width=0.95\linewidth]{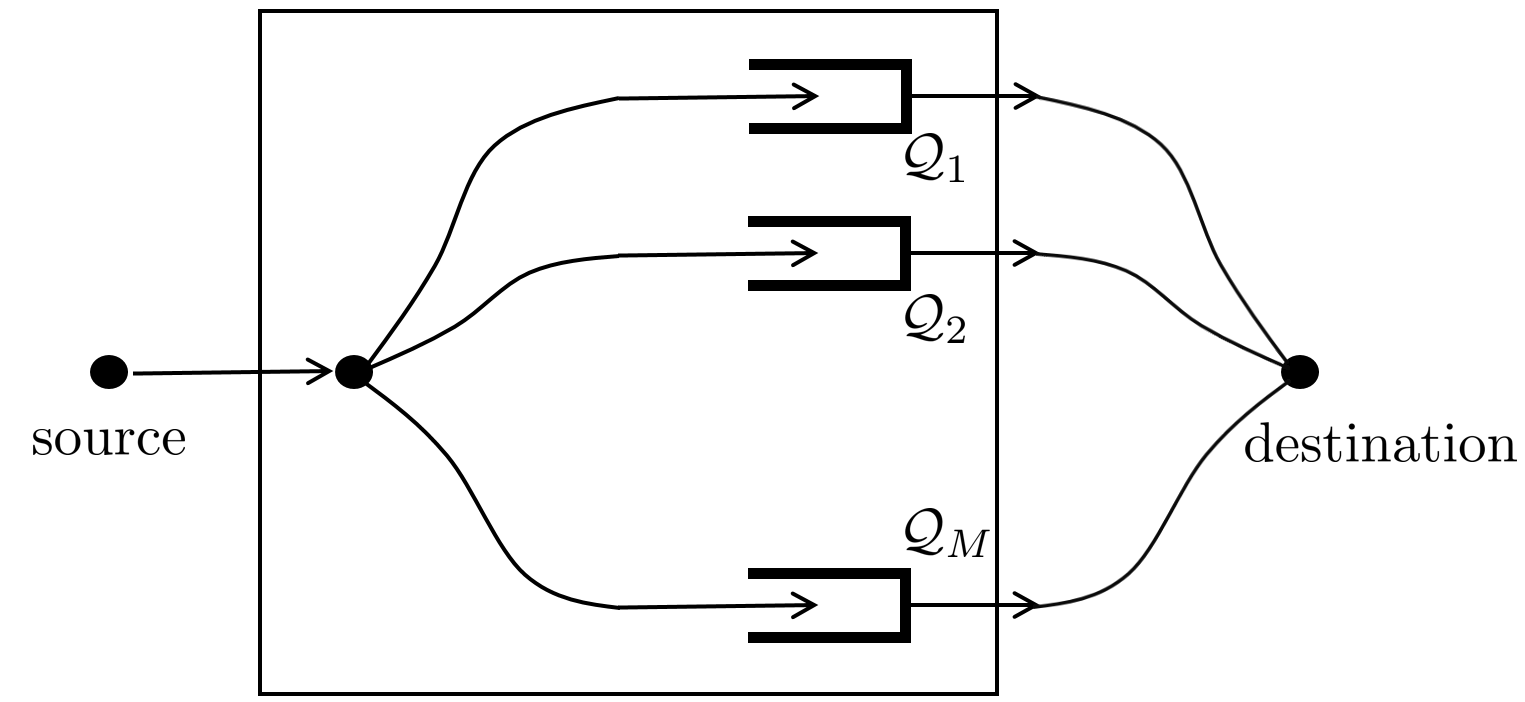}
  \caption{Illustration of the $M$ server queueing system.}\label{fig:queueing_system}
\end{figure}
A source generates update packets according to a renewal process, at a given rate $\lambda$. The update packets enter a queueing system, which consists of $M$ servers shown in Figure~\ref{fig:queueing_system}.
Each server has a rate $\mu$, and can service at most one update packet at any given time. The service times are independent and identically distributed across update packets. A scheduler determines routing and scheduling of update packets, which upon service reach the destination. Our goal is to ensure minimum age and/or minimum delay at the destination.

%The system designer has control over the service time distribution $F_S$ and the packet scheduling cum routing.
%A scheduler determines the server that an update packet connects for service, and the order in which each server services the packets assigned to it.

The system designer has control over three things:
\begin{enumerate}
  \item \emph{service}: it can decide the service time distribution, given the mean service time of $1/\mu$;
  \item \emph{routing}: it can determine the server that an update packet connects for service; and
  \item \emph{scheduling}: it can decide the order in which the update packets get serviced at each server.
\end{enumerate}
A \emph{scheduler} implements the routing and scheduling policy. The scheduler is also not allowed to drop any packets. Further, in determining the order of service of generated packets, we assume that the scheduler is not privy to the service times of the individual packets.
We also assume that only the service time distribution can be set before hand by the system designer, and not the service times of individual packets.

We use $X$ to denote the inter-generation time of update packets with distribution $F_X$, and $S$ to denote the service time random variable, with distribution $F_S$. Note that $\EX{X} = 1/\lambda$ and $\EX{S} = 1/\mu$ is fixed. We assume that $\lambda < \mu$, i.e. there is enough serving capacity in the network to service the generated updates.%\footnote{Several of these modeling assumptions can be relaxed to generate a more complex queuing network. We leave the investigation of the age-delay tradeoff in a more general queueing network for our future work.}
We use Minimize or $\min$, instead of the technically correct $\inf$, for ease of presentation. We now define the latency measures of packet delay, delay variance, and average age.

\subsection{Delay and Age of Information}
\label{sec:delay_aoi_def}
Let the update packets be generated at times $t_1, t_2, \ldots$, and let the update packet $i$ reach the destination at time $t^{'}_{i}$. The update packets may not reach the destination in the same order as they were generated. In Figure~\ref{fig:age}, packet $3$ reaches the destination before packet $2$, i.e. $t^{'}_{3} < t^{'}_{2}$. Delay for the $i$th packet is $D_i = t^{'}_{i} - t_{i}$, and the packet delay for the system is given by
\begin{equation}\label{eq:delay}
D = \limsup_{N \rightarrow \infty} \EX{\frac{1}{N}\sum_{i=1}^{N}D_i},
\end{equation}
where the expectation is over the update generation, service times, and scheduling discipline.
We skip a formal definition, but will use the notation $\VarD$ to denote variance in packet delay. For the $M$ server queueing system considered, we note that $\VarD$ is lower-bounded by the variance in service time distribution $F_S$.

Age of a packet $i$ is defined as the time since it was generated: $A^{i}(t) = (t-t_{i})\mathbb{I}_{\{ t > t_i\}}$,
%\begin{equation}\label{eq:pkt_age}
%A^{i}(t) = (t-t_{i})\mathbb{I}_{\{ t > t_i\}},
%\end{equation}
which is $0$ by definition for time prior to its generation $t < t_i$.
Age of information at the destination node, at time $t$, is defined as the minimum age across all received packets up to time $t$:
\begin{equation}\label{eq:age_t}
A(t) = \min_{i \in \mathcal{P}(t)} A^{i}(t),
\end{equation}
where $\mathcal{P}(t) \subset \{1, 2, 3, \ldots \}$ denotes the set of packets received by the destination, up to time $t$.
Notice that AoI increases linearly, and drops only at the times of certain packet receptions: $t^{'}_{1}, t^{'}_{3}, t^{'}_{4}, \ldots$, but not $t^{'}_{2}$ in Figure~\ref{fig:age}. Such an age drop occurs only when an update packet with a lower age, than all packets received thus far, is received by the destination.
We refer to such packets, that result in age drops, as the  \emph{informative packets}~\cite{2014ISIT_KamKomEp}.

We consider a time averaged metrics of age of information, namely, the average age. The average age is defined to be the time averaged area under the age curve:
\begin{equation}\label{eq:Aave}
A^{\text{ave}} = \limsup_{T \rightarrow \infty} \EX{\frac{1}{T}\int_{0}^{T}A(t) dt},
\end{equation}
where the expectation is over the packet generation and packet service processes.

It is important to note that the age $A(t)$, and therefore the average age, is defined from the view of the destination, and not a packet. $A(t)$ is the time since the last received informative packet was generated at the source. It, therefore, does not matter how long the non-informative packets take to reach the destination. This is unlike packet delay, which accounts for every packet in the system equally.

We use the notation $D(F_S, \pi_Q)$, $\VarD(F_S, \pi_Q)$, and $A^{\text{ave}}(F_S, \pi_Q)$ to make explicit the dependency of delay, its variance, and average age on the service time distribution $F_S$ and the queue scheduling policy $\pi_Q$.

In the next section, we pose the age-delay tradeoff problems. Age-delay tradeoff results are then proved for two special cases in Section~\ref{sec:lcfs} and Section~\ref{sec:inf_serv}, before arriving at the result for the $M$ server system in Section~\ref{sec:M_ServSyst}. %Results developed in Sections~\ref{sec:lcfs} and~\ref{sec:inf_serv}, help us prove those in Section~\ref{sec:M_ServSyst}.

%In the next section, we pose the age-delay tradeoff problems and also give the main results of this work. We show that the picture shown in Figure~\ref{fig:AoI_Delay_tradeoff1_plot1} holds true for the $M$-server queue. The proof of this, however, requires us to prove the age-delay tradeoff of various sub-problems, which we describe and prove in later sections.

\section{Age-Delay Tradeoff}
\label{sec:age_delay_tradeoff}

%\subsection{Age-Delay Tradeoff Problems}
%\label{sec:tradeoff_prob_def}

Motivated by the example in Figure~\ref{fig:AoI_Delay_tradeoff1_plot1}, we define two age-delay tradeoff problems. One, minimizes delay while the other minimizes delay variance, both over an average age constraint.
The age-delay tradeoff is defined as:
\begin{align}\label{eq:Aave_Delay_Tradeoff}
\begin{aligned}
T(\AoI) &= \underset{F_S, \pi_{Q}}{\text{Minimize}}
& & D(F_S, \pi_Q) \\
& \text{subject to} & & A^{\text{ave}}(F_S, \pi_Q) \leq \AoI, \\
& & & \EX{S} = 1/\mu.
\end{aligned}
\end{align}
Here, the function $T(\AoI)$ denotes the minimum delay that can be achieved for the $M$ server queueing system, with an average age constraint of $A^{\text{ave}}(F_S, \pi_Q) \leq \AoI$. It might seem that both minimum age and delay could be attained simultaneously. We will show that, $T(\AoI) \rightarrow \infty$ as $\AoI$ approaches the minimum average age
\begin{align}\label{eq:Amin}
\begin{aligned}
A_{\min} &= \underset{F_S, \pi_{Q}}{\text{Minimize}}
& & A^{\text{ave}}(F_S, \pi_Q), \\
& \text{subject to} & & \EX{S} = 1/\mu.
\end{aligned}
\end{align}
%In~\cite{talak18_determinacy}, we proved such a result for the LCFS queues with preemptive service (LCFSp). In this work, we show that such a result holds, even when the system designer has an option of choosing the queue scheduling discipline $\pi_Q$. This does not follow trivially from the LCFSp result in~\cite{talak18_determinacy}, primarily because LCFSp is not known to be the optimal scheduling discipline for single server systems, especially when the service times are not exponentially distributed~\cite{BedewyISIT17_LIFO_opt, sun_lcfs_better}.
%
%Our results imply that the minimum average age can be attained, but only at the cost of infinite packet delay.

Variability in packet delay is also an important metric in system performance. We define the age-delay variance tradeoff problem to be:
\begin{align}\label{eq:Aave_DelayVar_Tradeoff}
\begin{aligned}
V(\AoI) &= \underset{F_S, \pi_{Q}}{\text{Minimize}}
& & \VarD(F_S, \pi_Q) \\
& \text{subject to} & & A^{\text{ave}}(F_S, \pi_Q) \leq \AoI, \\
& & & \EX{S} = 1/\mu.
\end{aligned}
\end{align}
Here, the function $V(\AoI)$ denotes the minimum delay variance that can be achieved for the $M$ server queueing system, with an average age constraint of $A^{\text{ave}}(F_S, \pi_Q) \leq \AoI$. Counter to our intuition, we show that $V(\AoI) \rightarrow +\infty$ as \AoI~approaches its minimum value $A_{\min}$.

Ideally, we would like to obtain every point on the tradeoff curves, i.e., completely characterize the functions $T(\AoI)$ and $V(\AoI)$.
%This will tell us optimal queue scheduling policy and the service time distribution, for different requirements of average age.
The following result shows that the tradeoff curves can be done by optimizing a linear combination of average age and packet delay.
\begin{framed}
\begin{theorem}
\label{thm:aoi_delay_min}
The points on the age-delay tradeoff curve $(\AoI, T(\AoI))$ can be obtained by solving
\begin{align}\label{eq:aoi_delay_min}
\begin{aligned}
& \underset{F_S, \pi_{Q}}{\text{Minimize}}
& & D(F_S, \pi_Q) + \nu A^{\text{ave}}(F_S, \pi_Q)\\
& \text{subject to} & & \EX{S} = 1/\mu,
\end{aligned}
\end{align}
for all $\nu \geq 0$. Similarly, the points on the age-delay variance tradeoff curve $(\AoI, V(\AoI))$ can be obtained by solving~\eqref{eq:aoi_delay_min}, by replacing $D(F_S, \pi_Q)$ with $\VarD(F_S, \pi_Q)$.
\end{theorem}
\end{framed}
\begin{IEEEproof}
%The proof uses simple duality arguments, and is omitted due to space constraints.
This follows from Theorem II.2 in~\cite{Neely_NetworkOpt}.
\end{IEEEproof}
Theorem~\ref{thm:aoi_delay_min} motivates optimization of a latency metric that is a linear combination of average age and packet delay (or packet delay variance). This problem, however, is not easy to solve. For instance, in the case of a singe server, i.e. $M = 1$, with Poisson arrivals, the delay is minimized with deterministic service times and the variance in delay is minimized under the FCFS service discipline~\cite{1962_Kingman_Q_Discipline_Wait_Variance}. Exactly the opposite is true for the metric of average age. We will show in Section~\ref{sec:lcfs} that the LCFS queue scheduling policy with heavy tailed service minimizes average age. It, therefore, appears that the delay term and the average age term in~\eqref{eq:aoi_delay_min} are pulling the decision variables in opposite directions.
%
%We leave solving~\eqref{eq:aoi_delay_min}, and characterizing the age-delay tradeoff curve, for all values of $\AoI$, open for future investigation.

\begin{figure}
  \centering
  \includegraphics[width=0.75\linewidth]{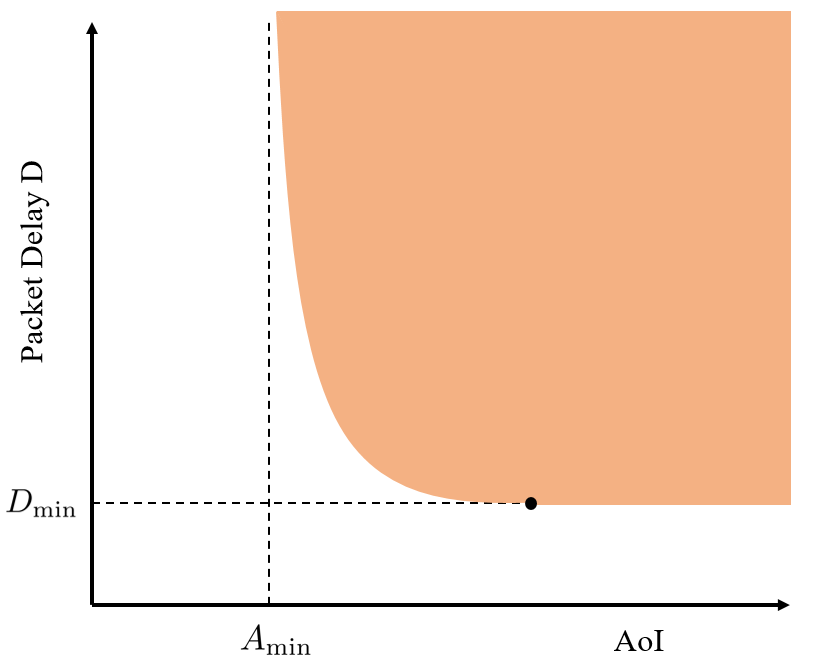}
  \caption{Illustration of strong age-delay tradeoff.}
  \label{fig:traeoff_illustration}
\end{figure}

%In the following sections, we will prove a that there is a strong age-delay tradeoff.
%
We say that a \emph{strong age-delay tradeoff} exists for $T(\AoI)$ if $T(\AoI) \rightarrow +\infty$ and $\AoI$ approaches $A_{\min}$. Conversely, \emph{no age-delay tradeoff} exists for  $T(\AoI)$ if the minimum average age and the minimum packet delay can be achieved simultaneously. Similar definition apply for age-delay variance tradeoff $V(\AoI)$.
Figure~\ref{fig:traeoff_illustration} illustrates a strong age-delay tradeoff. Note that this matches with our numerical results in Figure~\ref{fig:AoI_Delay_tradeoff1_plot1}.

We show that for the $M$ server system defined above there is a strong age-delay tradeoff when the update generation is Poisson, and a strong age-delay variance tradeoff for the general update generation process.
\begin{framed}
\begin{theorem}
\label{thm:M_tradeoff}
For the $M$ server system, the following statements are true:
\begin{enumerate}
  \item The minimum achievable average age is $A_{\min} = \frac{1}{2}\frac{\EX{X^2}}{\EX{X}}$.
  \item When update generation is Poisson, there is a strong age-delay tradeoff.
  \item When the update generation is a general renewal process, there is a strong age-delay variance tradeoff.
\end{enumerate}
\end{theorem}
\end{framed}

In Theorem~\ref{thm:M_tradeoff}, the update generation process is kept fixed. Thus, the strong age-delay tradeoffs hold even if we could control the inter-generation time distribution $F_X$, with a mean budget constraint of $\EX{X} = 1/\lambda$. The optimal update generation, with the mean constraint $\EX{X} = 1/\lambda$, that minimizes the $A_{\min}$ is the periodic update generation.

It seems counterintuitive at first that a strong tradeoff should exist between delay, or delay variance, and average age. However, a close examination reveals the following insight:

\emph{For age minimization it becomes necessary that the informative packets, the packets that reduce age, get serviced as soon as they arrive, while the non-informative packets, may incur as long a service time and queueing delay, as they do not matter in the age calculations. As we do this, the packet delay gets dominated by the delay of the non-informative packets, resulting in the two age-delay tradeoffs.}

In what follows, we prove this strong tradeoff between age-delay and age-delay variance. The proof of Theorem~\ref{thm:M_tradeoff} is given in Section~\ref{sec:M_ServSyst}. It relies on age-delay tradeoff results in two special cases, which are first studied in Sections~\ref{sec:lcfs} and~\ref{sec:inf_serv}.

In Section~\ref{sec:lcfs}, we consider the special case of a single server system, i.e. $M = 1$, in which the scheduling policy $\pi_Q$ is fixed to the last come first server with preemptive service (LCFSp). We prove the statements of Theorem~\ref{thm:M_tradeoff} for this special case. Namely, we we show that a age-delay tradeoff exists when the update generation is Poisson, and a age-delay variance tradeoff exists when the updates are generated according to a general renewal process.

In Section~\ref{sec:inf_serv}, we consider the special case of an infinite server queue, i.e. $M = \infty$, in which the scheduling policy $\pi_Q$ assigns every newly generate update to a new server. We again prove the statements of Theorem~\ref{thm:M_tradeoff} for this case.

Our approach in both these sections is as follows: We first derive an expression for the average age, and use it to obtain the minimum average age $A_{\min}$. We then use this to prove the two strong age-delay tradeoffs.
Using these two special cases, in Section~\ref{sec:M_ServSyst}, we finally prove Theorem~\ref{thm:M_tradeoff}.

\section{LCFSp Queue}
\label{sec:lcfs}
In this section, we consider a special case of the $M$ server system. We consider a single server system, i.e. $M = 1$, in which the queue scheduling discipline $\pi_Q$ is fixed to the LCFSp. The age and delay metrics, therefore, are just a function of the service time distribution $F_S$. The age-delay and the age-delay variance problem, for this case, reduces to
\begin{align}\label{eq:LCFS_Aave_Delay_Tradeoff}
\begin{aligned}
T(\AoI) &= \underset{F_S}{\text{Minimize}}
& & D(F_S) \\
& \text{subject to} & & A^{\text{ave}}_{\text{G/G/1}}(F_S) \leq \AoI, \\
& & & \EX{S} = 1/\mu,
\end{aligned}
\end{align}
and
\begin{align}\label{eq:LCFS_Aave_DelayVar_Tradeoff}
\begin{aligned}
V(\AoI) &= \underset{F_S}{\text{Minimize}}
& & \VarD(F_S) \\
& \text{subject to} & & A^{\text{ave}}_{\text{G/G/1}}(F_S) \leq \AoI, \\
& & & \EX{S} = 1/\mu,
\end{aligned}
\end{align}
where we use the notation $A^{\text{ave}}_{\text{G/G/1}}(F_S)$ to denote the average age for the LCFSp queue. The optimization is only over the service time distribution. We omit the dependence on $F_S$, whenever convenient.

The rest of this section is organized as follows. In Section~\ref{sec:lcfs_min_age}, we derive an expression for average age, and characterize the minimum average age. We also show that heavy tailed service time distributions achieve the minimum average age. In Section~\ref{sec:lcfs_age_delay_tradeoff}, we then use these results to prove that there is a strong age-delay and age-delay variance tradeoff. In Section~\ref{sec:lcfs_age_delay_disparity} point out a distinct age-delay disparity when the update generation is Poisson.

\subsection{Minimizing Age}
\label{sec:lcfs_min_age}

We first derive explicit expression average age for general inter-generation and service time distributions. We assume at least one of the distributions $F_{X}$ and $F_{S}$ to be continuous.

\begin{framed}
\begin{lemma}
\label{lem:LCFS_gg1}
The average age $A^{\text{ave}}_{\text{G/G/1}}(F_S)$ is given by
\begin{equation}\nonumber
A^{\text{ave}}_{\text{G/G/1}}(F_S) = \frac{1}{2}\frac{\EX{X^2}}{\EX{X}} + \frac{\EX{\min\left(X, S\right) }}{\pr{S < X}},
\end{equation}
where $X$ and $S$ denotes the independent inter-generation and service time distributed random variables, respectively.
\end{lemma}
\end{framed}
\begin{IEEEproof}
\begin{figure}
  \centering
  \includegraphics[width=\linewidth]{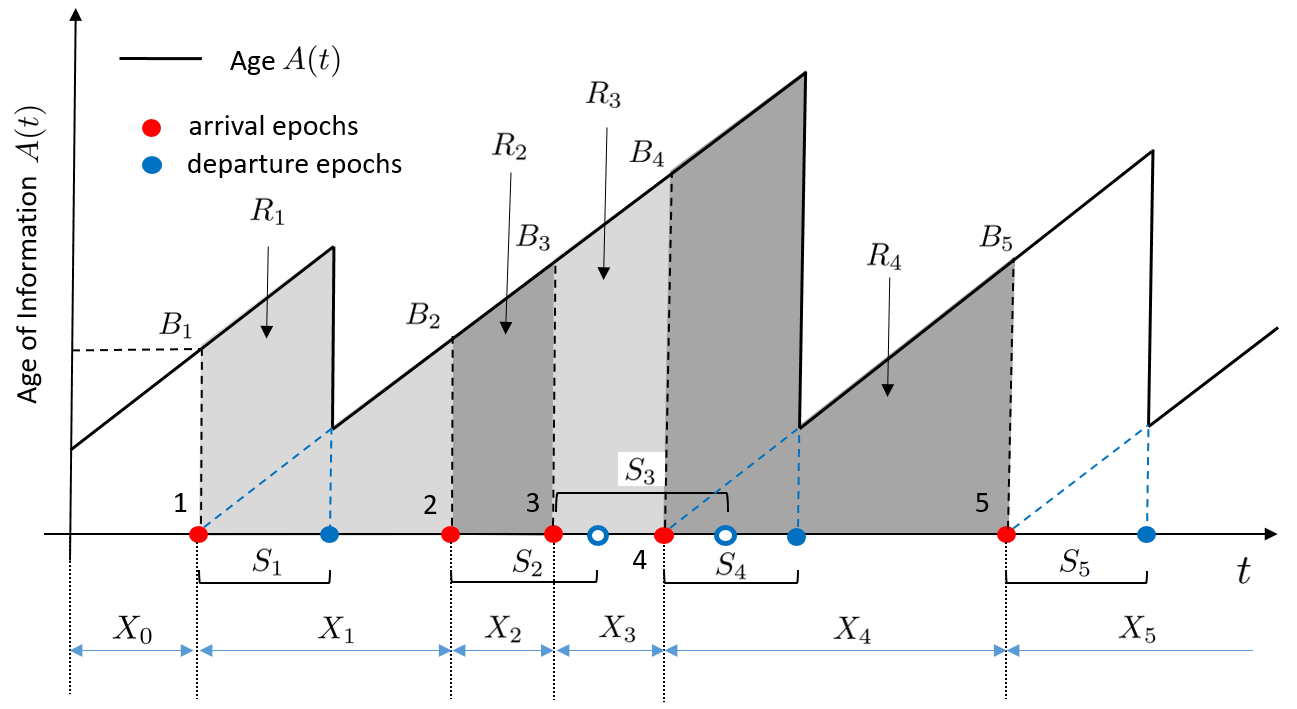}
  \caption{Age $A(t)$ evolution in time $t$ for the LCFS queue with preemptive service.}\label{fig:lcfs}
\end{figure}
Let $X_i$ denote the inter-generation time between the $i$th and $(i+1)$th update packet. Due to preemption, not all packets get serviced on time to contribute to age reduction. We illustrate this in Figure~\ref{fig:lcfs}. Observe that packets $2$ and $3$ arrive before packet $4$. However, packet $2$ is preempted by packet $3$, which is subsequently preempted by packet $4$. Thus, packet $4$ is serviced before $2$ and $3$. Service of packet $2$ and $3$ (not shown in figure) does not contribute to age curve $A(t)$ because they contain stale information.

In order to analyze this, define $S_{i}$ to be the virtual service time for packet $i$, such that $\{ S_i \}_{i \geq 1}$ are i.i.d., and distributed according to the service time distribution $F_{S}$. If $S_{i} < X_{i}$, then the packet $i$ is serviced, and the age $A(t)$ drops to $S_{i}$, which is the time since generation of the packet $i$. In Figure~\ref{fig:lcfs}, we observe this for packets $1$, $4$ and $5$. However, if $S_{i} > X_{i}$, the service of packet $i$ is preempted, and the server starts serving the newly arrived packet $(i+1)$. In Figure~\ref{fig:lcfs}, observe that $S_2 > X_2$ and $S_3 > X_3$, while $S_4 < X_4$, and thus, packet $4$ gets serviced before $2$ and $3$.

For computing the average age, which is nothing but the time averaged area under the age curve $A(t)$, we compute the sum $\sum_{i=1}^{M} R_i$, where $R_i$ is the area under $A(t)$ between the $i$th and $(i+1)$th generation of update packets; see Figure~\ref{fig:lcfs}. To do so, we obtain a recursion for $B_i$, the age $A(t)$ at the time of generation of the $i$th update packet: define $Z_i \triangleq \sum_{k=0}^{i-1} X_k$ and $B_i = A(Z_i)$, and show that
\begin{equation}
R_i = \left\{ \begin{array}{cc}
                B_i X_i + \frac{1}{2}X_{i}^{2} & \text{if}~X_i < S_i \\
                B_i S_i + \frac{1}{2}X_{i}^{2} & \text{if}~X_i \geq S_i
              \end{array}\right..
\end{equation}
The detailed proof is given in Appendix~\ref{pf:lem:LCFS_gg1}.
\end{IEEEproof}

\begin{figure}
  \centering
  \includegraphics[width=\linewidth]{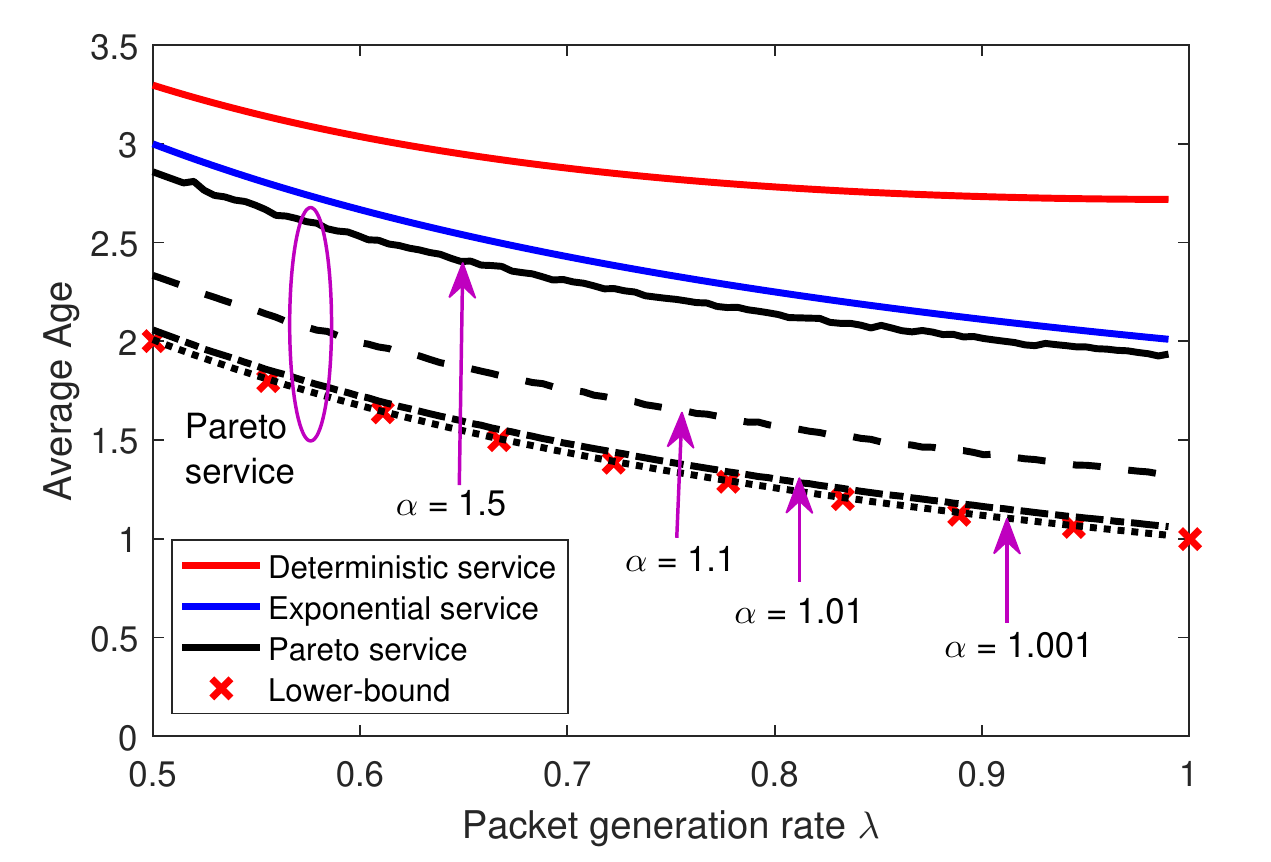}
  \caption{Plotted is the average age under deterministic, exponential, and Pareto ($\alpha = 1.5, 1.1, 1.01,$ and $1.001$) distributed service times distributions for the LCFS queue with preemptive service. Service rate $\mu = 1$, while the packet generation rate $\lambda$ varies from $0.5$ to $0.99$.}
  \label{fig:LCFS_MG1_ave_Par}
\end{figure}
We now prove that a heavy tailed continuous service time distribution minimizes the average age. In Figure~\ref{fig:LCFS_MG1_ave_Par}, we plot average age as a function of packet generation rates $\lambda$, for three different service time distributions: deterministic service, exponential service, and Pareto service. The cumulative distribution function for a Pareto service distribution, with mean $1/\mu$, is given by
\begin{equation}\label{eq:Par_Dist}
F_{S}(s) = \left\{ \begin{array}{cc}
                     1 - \left( \frac{\theta(\alpha)}{s}\right)^{\alpha} &~\text{if}~s \geq \theta(\alpha) \\
                     0 &~\text{otherwise}
                   \end{array}\right.,
\end{equation}
where $\theta(\alpha) = \frac{1}{\mu}\left( 1 - \frac{1}{\alpha}\right)$ and $\alpha > 1$ is the shape parameter. The shape parameter $\alpha$ determines the tail of the distribution. The closer the shape parameter is to $1$, the heavier is the tail.

We observe in Figure~\ref{fig:LCFS_MG1_ave_Par} that the Pareto service yields better age than the exponential service. Furthermore, observe that the heavier the tail of the Pareto distribution, i.e. the closer $\alpha$ is to $1$, the lower is the age.
Also plotted is the age lower-bound $1/\lambda$, as no matter what the service, the age cannot decrease below the inverse rate at which packets are generated.

\begin{figure}
  \centering
  \includegraphics[width=\linewidth]{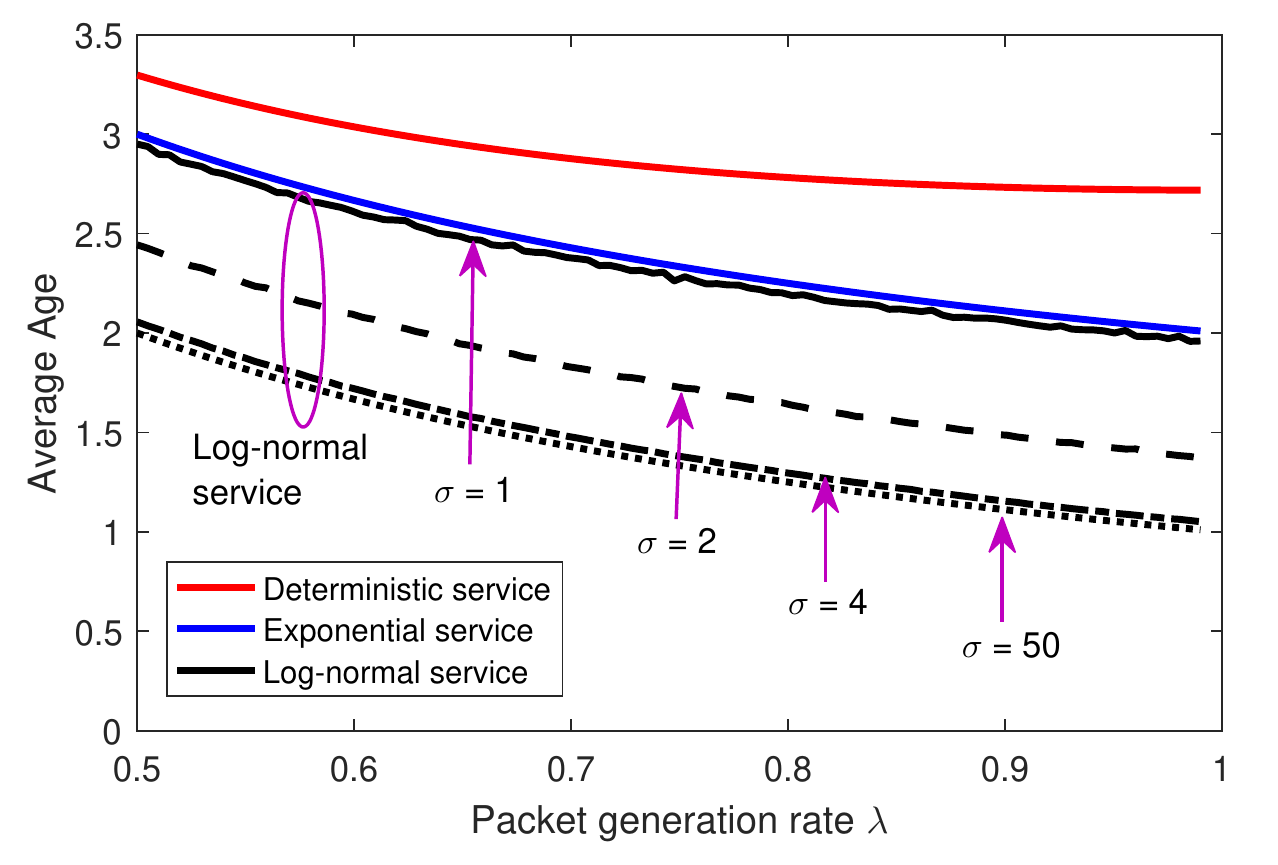}
  \caption{Plotted is the average age under deterministic, exponential, and log-normal ($\sigma = 1, 2, 4,$ and $50$) distributed service times distributions for the LCFS queue with preemptive service. Service rate $\mu = 1$, while the packet generation rate $\lambda$ varies from $0.5$ to $0.99$.}
  \label{fig:LCFS_MG1_ave_LN}
\end{figure}

We observe similar behavior not just for Pareto distributed service, but also for other heavy tailed distributions. In Figure~\ref{fig:LCFS_MG1_ave_LN}, we plot average age for log-normal service distribution, another heavy-tail distribution, with mean $1/\mu$ given by:
\begin{equation}\label{eq:log_normal}
S = \exp\left\{ -\log\mu - \frac{\sigma^2}{2} + \sigma N\right\},
\end{equation}
where $N \sim \mathcal{N}(0,1)$ is the standard normal distribution and $\sigma$ is a parameter that determines the tail of the distribution $F_S$. Higher $\sigma$ implies heavier tail, and in Figure~\ref{fig:LCFS_MG1_ave_LN} we observe that it results in smaller age, that approaches the age lower-bound of $1/\lambda$ as $\sigma \rightarrow +\infty$. We observe similar behavior for Weibull distributed service, with mean $1/\mu$:
\begin{equation}\label{eq:Weibull}
F_{S}(s) = 1 - e^{-\left( s /\beta \right)^{\kappa}},
\end{equation}
for all $s \geq 0$, where $\beta = \left[ \mu \Gamma(1 + 1/\kappa) \right]^{-1}$, as $\kappa \downarrow 0$; here $\Gamma(x) = \int_{0}^{\infty} t^{x-1} e^{-t} dt$ is the gamma function.

We now show that the minimum average age is given by $A_{\min} = \frac{1}{2}\frac{\EX{X^2}}{\EX{X}}$. We first prove that the average age is lower-bounded by $\frac{1}{2}\frac{\EX{X^2}}{\EX{X}}$ for any service time distribution, and then show that this lower-bound is in fact achievable for the three heavy tailed service time distributions.
\begin{framed}
\begin{theorem}
\label{thm:LCFS_GG1_heavy_tail_opt}
The average age is lower bounded by
\begin{equation}\nonumber
A^{\text{ave}}_{\text{G/G/1}}(F_S) \geq \frac{1}{2}\frac{\EX{X^2}}{\EX{X}}.
\end{equation}
Further, this lower-bound is achieved for
\begin{enumerate}
  \item Pareto distributed service~\eqref{eq:Par_Dist} as $\alpha \rightarrow 1$,
  \item Log-normal distributed service~\eqref{eq:log_normal} as $\sigma \rightarrow +\infty$, and
  \item Weibull distributed service~\eqref{eq:Weibull} as $\kappa \rightarrow 0$.
\end{enumerate}
\end{theorem}
\end{framed}
\begin{IEEEproof}
The lower-bound on the average age follows directly from the age expressions obtained in Lemma~\ref{lem:LCFS_gg1}, and noticing that $\frac{\EX{\min\{ X, S \}}}{\pr{S < X}} \geq 0$. The distributions, namely the Pareto, log-normal, and Weibull, are all parametric distributions parameterized here by $\alpha$, $\sigma$, and $\kappa$, respectively. We, therefore, prove the following generic result, which gives us a sufficient conditions for the optimality of the average age for a general, parametric continuous service time distribution $F_{S}$, parameterized by $\eta$. We hide the dependence of the parameter $\eta$ on $S$ and $F_S$ for notational convenience.
\begin{framed}
\begin{lemma}
\label{lem:suff_cond_lcfs}
Let a parametric, continuous service time $S$, with parameter $\eta$, satisfy
\begin{enumerate}
  \item $\EX{S} = 1/\mu$ for all $\eta$,
  \item $\EX{\mathbb{I}_{\{ S > x\}}} \rightarrow 0$ as $\eta \rightarrow \eta^{\ast}$, and
  \item $\EX{S\mathbb{I}_{\{ S \leq x\}}} \rightarrow 0$ as $\eta \rightarrow \eta^{\ast}$,
\end{enumerate}
for all $x > 0$ and for some $\eta^{\ast}$. Then,
\begin{equation}
\lim_{\eta \rightarrow \eta^{\ast}} A^{\text{ave}}_{\text{G/G/1}}(F_{S}) = \frac{1}{2}\frac{\EX{X^2}}{\EX{X}}.
\end{equation}
\end{lemma}
\end{framed}
\begin{IEEEproof}
Let for a parametric, continuous service time distribution $F_{S}$ the stated properties hold.
Note that
\begin{equation}\nonumber
\EX{\min\{X, S\}} = \EX{X\mathbb{I}_{\{ S \geq X\}}} + \EX{S\mathbb{I}_{\{S < X \}}},
\end{equation}
Using conditions 2 and 3 in the Lemma, and the bounded convergence theorem~\cite{Durrett}, we have $\EX{X\mathbb{I}_{\{ S \geq X\}}} \rightarrow 0$, $\EX{S \mathbb{I}_{\{S  < X \}}} \rightarrow 0$, and $\pr{S < X} \rightarrow 1$ as $\eta \rightarrow \eta^{\ast}$. Substituting all this in the average age expression in Lemma~\ref{lem:LCFS_gg1}, we obtain $A^{\text{ave}}_{\text{G/G/1}}(F_{S}) \rightarrow \frac{1}{2}\frac{\EX{X^2}}{\EX{X}}$ as $\eta \rightarrow \eta^{\ast}$.
\end{IEEEproof}

It, therefore, suffices to prove that the sufficient conditions in Lemma~\ref{lem:suff_cond_lcfs} are satisfied by the Pareto, log-normal, and Weibull distributions. We know, by definition, that all these distributions are continuous and have mean $\EX{S} = 1/\mu$ for all parameter values. The other conditions are verified in Appendix~\ref{pf:heavy_tail}.
\end{IEEEproof}

We showed that the minimum average age $A_{\min} = \frac{1}{2}\frac{\EX{X^2}}{\EX{X}}$ is achieved under the three heavy tailed service time distributions. Such heavy tailed services' induce maximum variation in the service times, and thus, will yield a worse delay and delay variance. However, it is not clear whether these are the only distributions that can achieve minimum age. Perhaps, we may be able to find a distribution $F_S$, that minimizes age and well as delay and delay variance. In the next section, we prove that this is not so, and that there is a strong age-delay and age-delay variance tradeoff.

\subsection{Age-Delay Tradeoff}
\label{sec:lcfs_age_delay_tradeoff}

We now prove that there exists a strong age-delay and age-delay variance tradeoff for the single server system, when the queue scheduling is fixed to LCFSp.
\begin{framed}
\begin{theorem}
For a single server system under LCFSp scheduling policy, the following statements are true:
\begin{enumerate}
  \item When the update generation is Poisson, there is a strong age-delay tradeoff.
  \item When the update generation is a general renewal process, there is a strong age-delay variance tradeoff.
\end{enumerate}
\end{theorem}
\end{framed}
\begin{IEEEproof}
Let $A_{\min} = \frac{1}{2}\frac{\EX{X^2}}{\EX{X}}$ denote the minimum average age. We have to show that as $\AoI \rightarrow A_{\min}$ in~\eqref{eq:LCFS_Aave_Delay_Tradeoff}, $T(\AoI) \rightarrow +\infty$, when the updates are generated according to a Poisson process. We also have to show that as $\AoI \rightarrow A_{\min}$ in~\eqref{eq:LCFS_Aave_DelayVar_Tradeoff}, $V(\AoI) \rightarrow +\infty$ for general update generation process.

When the update generation is Poisson, the queue is a M/G/1 LCFSp queue. The packet delay for this queue is given by~\cite{data_nets}:
\begin{equation}
D(F_S) = \frac{\lambda}{2}\frac{\EX{S^2}}{1-\rho} + \EX{S}.
\end{equation}
Furthermore, the delay variance is lower-bounded by the variance in service time, namely, $\text{VarD}(F_S) \geq \EX{S^2} - \EX{S}^2$.
Therefore, it suffices to show that as $\AoI \rightarrow A_{\min}$ in~\eqref{eq:LCFS_Aave_Delay_Tradeoff} and~\eqref{eq:LCFS_Aave_DelayVar_Tradeoff} we have $\EX{S^2} \rightarrow +\infty$.

In the following, we prove the strong age-delay tradeoff. The arguments are exactly the same for establishing the strong age-delay variance tradeoff, as we only have to show that $\EX{S^2} \rightarrow +\infty$.

To establish the strong age-delay tradeoff, we use the expressions for average age derived in Lemma~\ref{lem:LCFS_gg1}. Let $S_{\AoI}$ denote the service time, and $F_{S_{\AoI}}$ the corresponding service time distribution, that solves~\eqref{eq:LCFS_Aave_Delay_Tradeoff}. Now, as $\AoI \rightarrow A_{\min}$ in~\eqref{eq:LCFS_Aave_Delay_Tradeoff} we must have
\begin{multline}
A^{\text{ave}}_{\text{G/G/1}}(F_{S_{\AoI}}) = \frac{1}{2}\frac{\EX{X^2}}{\EX{X}} + \frac{\EX{\min\left(X, S_{\AoI}\right) }}{\pr{S_{\AoI} < X}} \\
\rightarrow \frac{1}{2}\frac{\EX{X^2}}{\EX{X}} = A_{\min},
\end{multline}
which implies
\begin{equation}
\label{eq:nuance1}
\lim_{\AoI \rightarrow A_{\min}} \frac{\EX{\min\left(X, S_{\AoI}\right) }}{\pr{S_{\AoI} < X}} = 0.
\end{equation}
Now, notice that $\pr{S_{\AoI} < X}$, being probability, is bounded by $1$. Therefore, for~\eqref{eq:nuance1} to hold, it must be the case that
\begin{equation}\label{eq:nuance2}
\lim_{\AoI \rightarrow A_{\min}} \EX{\min\left(X, S_{\AoI}\right)} = 0.
\end{equation}

Substituting the fact $\EX{\min\left(X, S_{\AoI}\right)} = \EX{X\mathbb{I}_{\{X < S_{\AoI}\}}} + \EX{S_{\AoI} \mathbb{I}_{\{ X \geq S_{\AoI}\}}}$ in~\eqref{eq:nuance2} we get
\begin{equation}
\label{eq:nuance3}
\lim_{\AoI \rightarrow A_{\min}} \EX{X\mathbb{I}_{\{X < S_{\AoI}\}}} = 0,
\end{equation}
and
\begin{equation}\label{eq:nuance3b}
\lim_{\AoI \rightarrow A_{\min}} \EX{S_{\AoI} \mathbb{I}_{\{ X \geq S_{\AoI}\}}} = 0.
\end{equation}
Now,~\eqref{eq:nuance3} and~\eqref{eq:nuance3b} implies that there exists a $x_0 > 0$ such that
\begin{equation}
\label{eq:nuance4}
\lim_{\AoI \rightarrow A_{\min}} \EX{\mathbb{I}_{\{x_0 < S_{\AoI}\}}} = 0,
\end{equation}
and
\begin{equation}
\label{eq:nuance4b}
\lim_{\AoI \rightarrow A_{\min}} \EX{S_{\AoI} \mathbb{I}_{\{ x_0 \geq S_{\AoI}\}}} = 0.
\end{equation}
This can be established by a short proof by contradiction.
Using Lemma~\ref{lem:exists_to_all} in Appendix~\ref{app:serv_dist_prop}, along with~\eqref{eq:nuance4} and~\eqref{eq:nuance4b}, we obtain
\begin{equation}
\label{eq:nuance5}
\lim_{\AoI \rightarrow A_{\min}} \EX{\mathbb{I}_{\{x < S_{\AoI}\}}} = 0,
\end{equation}
and
\begin{equation}\label{eq:nuance5b}
\lim_{\AoI \rightarrow A_{\min}} \EX{S_{\AoI} \mathbb{I}_{\{ x \geq S_{\AoI}\}}} = 0,
\end{equation}
for all $x \geq x_0$. Lemma~\ref{lem:C_implies_S2} in Appendix~\ref{app:serv_dist_prop} shows that these two conditions in~\eqref{eq:nuance5} and~\eqref{eq:nuance5b} imply
\begin{equation}
\lim_{\AoI \rightarrow A_{\min}} \EX{S_{\AoI}^2} = +\infty,
\end{equation}
which proves our result.
\end{IEEEproof}

In the next subsection, we bring out an even stronger contrast between age and delay, than the strong age-delay tradeoff. We show that the the delay minimizing service time distribution results in the worst case average age, and that the average age minimizing service time distribution results in the worst case delay.

\subsection{Age-Delay Disparity under Poisson Update Generation}
\label{sec:lcfs_age_delay_disparity}
To bring out the contrast between packet delay and AoI metrics, we consider the special case in which the update packets are generated according to a Poisson process. In this case, the single server system is nothing but a M/G/1 LCFSp queue. We now show that deterministic service yields the worst average age, across all service time distributions. We use the notation $A^{\text{ave}}_{G/G/1}$ for average age for the G/G/1 LCFSp queue, and omit the dependence on service time distribution $F_S$ for convenience.
\begin{framed}
\begin{theorem}
\label{thm:opt_LCFS_mg1}
For a single server system under LCFSp scheduling policy and Poisson update generation, the deterministic service yields the worst case average age:
\begin{align}
A^{\text{ave}}_{\text{M/G/1}} &\leq A^{\text{ave}}_{\text{M/D/1}}. \nonumber
\end{align}
\end{theorem}
\end{framed}
\begin{IEEEproof}
See Appendix~\ref{pf:thm:opt_LCFS_mg1}.
\end{IEEEproof}
It should be intuitive that if the packets in service are often preempted, then very few packets complete service on time, and this results in a very high AoI.
It turns out that deterministic service maximizes the probability of preemption. For the LCFS M/G/1 queue, the probability of preemption is given by $\pr{S > X} = 1 - \EX{e^{-\lambda S}}$, as $X$ is exponentially distributed with rate $\lambda$. This can be upper-bounded by $1 - e^{-\lambda\EX{S}} = \pr{\EX{S} > X}$, using Jensen's inequality, which is nothing but the probability of preemption under deterministic service: $S = \EX{S}$ almost surely.

Comparing age with packet delay for the LCFSp queue results in a peculiar conclusion. The packet delay for a M/G/1 LCFSp queue is given by~\cite{data_nets}:
\begin{equation}\nonumber
D = \frac{\lambda}{2}\frac{\EX{S^2}}{1-\rho} + \EX{S}.
\end{equation}
Note that this expression of packet delay $D$ is minimized when the service time $S$ is deterministic, namely $S = \EX{S}$ almost surely; follows from Jensen's inequality $\EX{S^2} \geq \EX{S}^2$. However, from Theorem~\ref{thm:opt_LCFS_mg1} we know that deterministic service time maximizes age.
This leads to the conclusion that, for the M/G/1 LCFSp queue, \emph{the service time distribution that minimizes delay, maximizes age of information}.
%
%It is also noteworthy that the three heavy tailed service time distributions, which minimize peak and average age, have $\EX{S^2} \rightarrow +\infty$, and therefore, result in unbounded packet delay.

\section{Infinite Servers}
\label{sec:inf_serv}
In this section, we consider the case when $M = \infty$, i.e. there are infinite servers in the system. The queue scheduling policy $\pi_Q$ is also fixed, and it assigns a new server to every arriving update packet. We call this the \emph{work conserving} scheduling policy. The infinite server system, under the work conserving policy, is nothing but the G/G/$\infty$ queue. With the scheduling policy $\pi_Q$ fixed, the age and delay metrics are just a function of the service time distribution $F_S$.

Note that under the above scheduling policy, the packet delay incurred equals the service time, and thus, $D = \EX{S} = 1/\mu$. This implies that the minimum age and minimum delay, which is $1/\mu$, can be achieved simultaneously, and the service time distribution that achieves this can be obtained by solving~\eqref{eq:Amin}.

The age-delay variance tradeoff problem, on the other hand, is not so trivial. This can be written as
\begin{align}\label{eq:Inf_Aave_DelayVar_Tradeoff}
\begin{aligned}
V(\AoI) &= \underset{F_S}{\text{Minimize}}
& & \VarD(F_S) \\
& \text{subject to} & & A^{\text{ave}}_{\text{G/G/}\infty}(F_S) \leq \AoI, \\
& & & \EX{S} = 1/\mu,
\end{aligned}
\end{align}
where the notation $A^{\text{ave}}_{\text{G/G/}\infty}(F_S)$ denotes the average age for the G/G/$\infty$ queue. The optimization is only over the service time distribution. We omit the dependence on $F_S$, whenever convenient.

The rest of this section is organized as follows. In Section~\ref{sec:inf_min_age}, we derive an expressions for the average age, and characterize its minimum. We also show that heavy tailed service time distributions achieve the minimum average age. In Section~\ref{sec:inf_age_delay_tradeoff}, we use these results to prove that there is a strong age-delay variance tradeoff. In Section~\ref{sec:inf_age_delay_disparity}, we obtain an average age maximizing service time distribution, and point to the disparity between the average age metric and delay variance.

\subsection{Minimizing Age}
\label{sec:inf_min_age}
We first derive an expression for average age for the system.
\begin{framed}
\begin{lemma}
\label{lem:gginf}
For the G/G/$\infty$ queue, the average is given by
%\begin{equation}
%\nonumber
%A^{\text{p}}_{\text{G/G/}\infty} = \EX{X} + \EX{\min_{l \geq 0}\left\{ \sum_{k=1}^{l}X_{k} + S_{l+1}\right\} },
%\end{equation}
%and
\begin{equation}
A^{\text{ave}}_{\text{G/G/}\infty}(F_S) = \frac{1}{2}\frac{\EX{X^2}}{\EX{X}} + \EX{\min_{l \geq 0}\left\{ \sum_{k=1}^{l}X_{k} + S_{l+1}\right\} }, \nonumber
\end{equation}
where $X$ and $\{ X_{k} \}_{k \geq 1}$ are i.i.d. distributed according to $F_X$, while $\{ S_{k} \}_{k \geq 1}$ are i.i.d. distributed according to $F_{S}$.
\end{lemma}
\end{framed}
\begin{IEEEproof}
For the G/G/$\infty$ queue, each arriving packet is serviced by a different server. As a result, the packets may get serviced in an out of order fashion. Figure~\ref{fig:gginf}, which plots age evolution for the G/G/$\infty$ queue, illustrates this. In Figure~\ref{fig:gginf}, observe that packet $3$ completes service before packet $2$. As a result, the age doesn't drop at the service of packet $3$, as it now contains stale information. To analyze average age, it is important to characterize these events of out of order service.
\begin{figure}
  \centering
  \includegraphics[width=\linewidth]{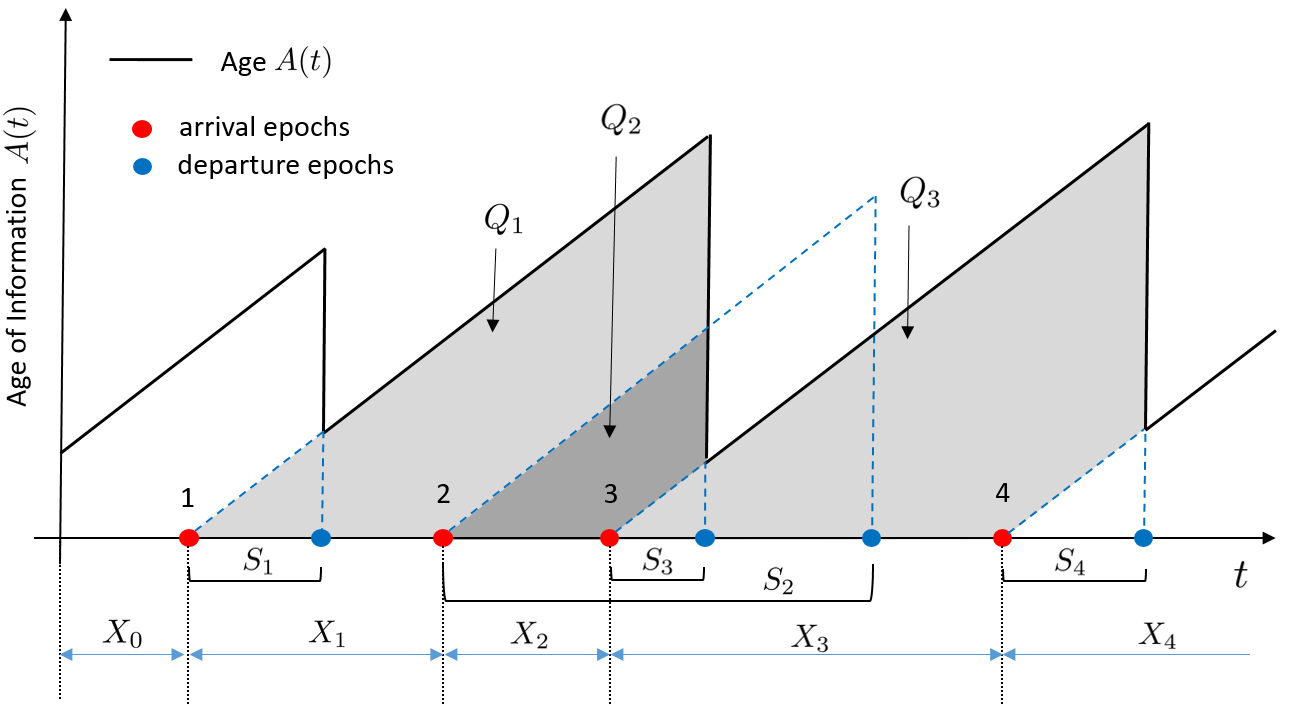}
  \caption{Age $A(t)$ evolution over time $t$ for G/G/$\infty$ queue.}
  \label{fig:gginf}
\end{figure}

Let $X_i$ denote the inter-generation time between the $i$th and $(i+1)$th packet, and $S_i$ denote the service time for the $i$th packet. In
Figure~\ref{fig:gginf}, $X_2 + S_3 < S_2$, and therefore, packet $3$ completes service before packet $2$. To completely characterize this,
define $Z_i \triangleq \sum_{k=0}^{i-1} X_k$ to be the time of generation of the $i$th packet.
Note that the $i$th packet gets serviced at time $Z_i + S_i$, the $(i+1)$th packet gets services at time $Z_i + X_i + S_{i+1}$, and similarly, the $(i+l)$th packet gets serviced at time $Z_i + \sum_{k=1}^{l}X_{i+k-1} + S_{i+l}$, for all $l \geq 1$.
Let $D_i$ denote the time from the $i$th packet generation to the time there is a service of the $i$th packet, or a packet that arrived after the $i$th packet, whichever comes first. Thus,
\begin{align}
D_i &= \min\{S_i, X_{i} + S_{i+1}, X_{i} + X_{i+1} + S_{i+2}, \ldots \} \nonumber \\
    &= \min_{l \geq 0}\left\{ \sum_{k=1}^{l}X_{i+k-1} + S_{i+l}\right\}.
\end{align}
In Figure~\ref{fig:gginf}, note that $D_1 = S_1$, $D_2 = X_2 + S_3$, $D_3 = S_3$, and $D_4 = S_4$.

%We note that the $i$th peak is nothing but $X_i + D_{i+1}$. Therefore, the peak age is given by
%\begin{align}\nonumber
%A^{\text{p}}_{\text{G/G/}\infty} &= \EX{X_i} + \EX{D_{i+1}}, \\
%&= \EX{X_i} + \EX{\min_{l \geq 0}\left\{ \sum_{k=1}^{l}X_{i+k-1} + S_{i+l}\right\}}, \nonumber \\
%&= \EX{X} + \EX{\min_{l \geq 0}\left\{ \sum_{k=1}^{l}X_{k-1} + S_{l}\right\}}, \nonumber
%\end{align}
%since the $X_i$s and $S_i$s are independent and identically distributed.

The area under the age curve $A(t)$ is nothing but the sum of the areas of the trapezoids $Q_i$ (see Figure~\ref{fig:gginf}). Applying the renewal reward theorem~\cite{wolff}, by letting the reward for the $i$th renewal, namely $[Z_i, Z_i + X_i)$, be the area $Q_i$, we get the average age to be:
\begin{equation}\label{eq:z0}
A^{\text{ave}}_{\text{G/G/}\infty}(F_S) = \frac{\EX{Q_i}}{\EX{X_i}}.
\end{equation}
It is easy to see that
\begin{equation}
Q_i = \frac{1}{2}(X_i + D_{i+1})^2 - \frac{1}{2}D^{2}_{i+1}, \label{eq:z1}
\end{equation}
as the trapezoid $Q_i$ extends from the time of the $i$th packet generation to the time at which the $(i+1)$th, or a packet that arrives after the $(i+1)$th packet, is served; which is nothing but $X_{i} + D_{i+1}$. For illustration, note that $Q_1 = \frac{1}{2}(X_1 + X_2 + S_3)^2 - \frac{1}{2}(X_2 + S_3)^2$, which is same as~\eqref{eq:z1}, for $i = 1$, since $D_2 = X_2 + S_3$. Substituting~\eqref{eq:z1} in~\eqref{eq:z0}, we obtain
\begin{equation}
A^{\text{ave}}_{\text{G/G/}\infty}(F_S) = \frac{1}{2}\frac{\EX{X^2}}{\EX{X}} + \frac{\EX{X_i D_{i+1}}}{\EX{X_i}}.
\end{equation}
We obtain the result by noting that $X_i$ and $D_{i+1}$ are independent.
%See Appendix~\ref{pf:lem:gginf}.
\end{IEEEproof}

\begin{figure}
  \centering
  \includegraphics[width=\linewidth]{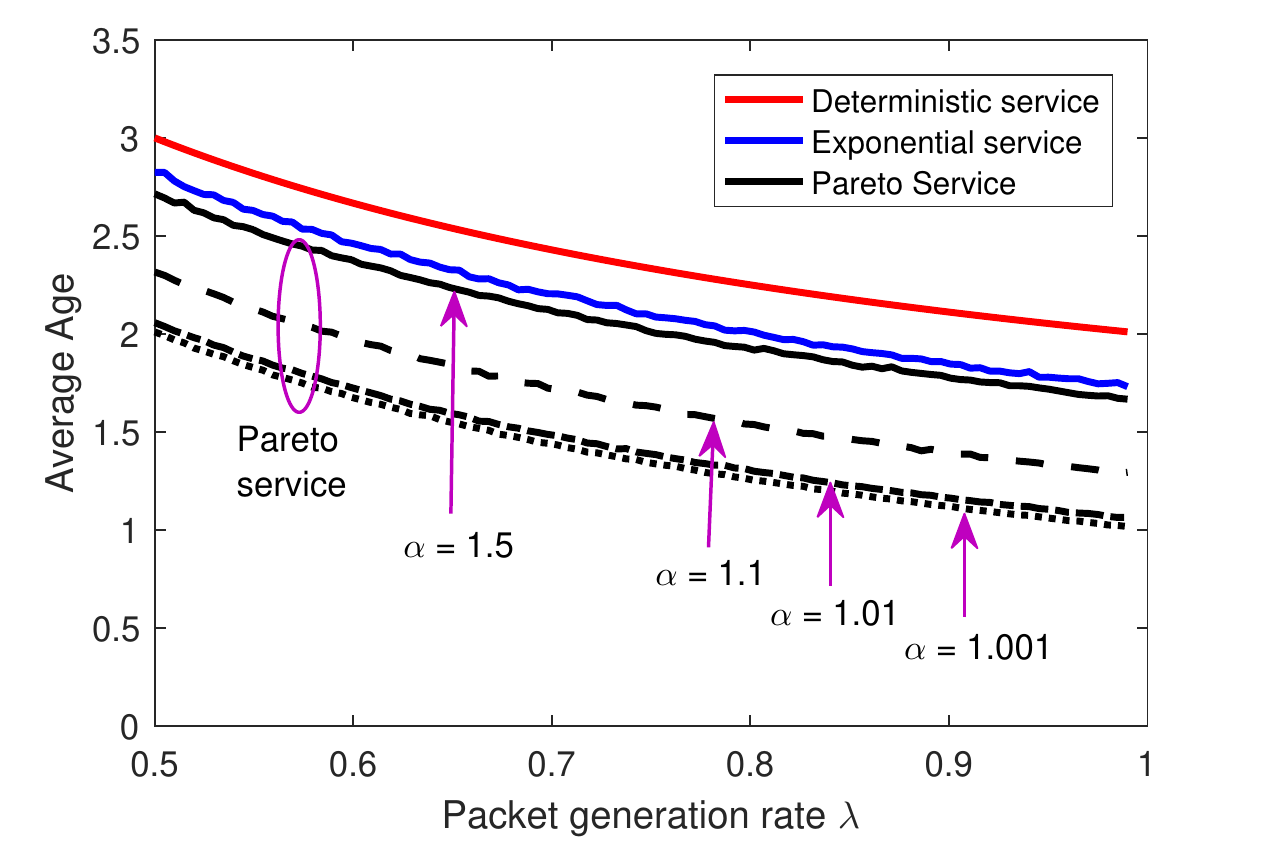}
  \caption{Plotted is the average age under deterministic, exponential, and Pareto ($\alpha = 1.5, 1.1, 1.01,$ and $1.001$) distributed service times distributions for the infinite server M/G/$\infty$ queue. Service rate $\mu = 1$, while the packet generation rate $\lambda$ varies from $0.5$ to $0.99$.}
  \label{fig:InfServ_ave_Par}
\end{figure}

In Figure~\ref{fig:InfServ_ave_Par}, we plot the average age for the M/G/$\infty$ queue under three service distributions: deterministic, exponential, and Pareto distribution (given in~\eqref{eq:Par_Dist}), with mean $1/\mu$. We observe that the heavy tail Pareto distributed service performs better than the exponential service. Also, heavier tail or decreasing $\alpha$ results in improvement in age. It appears, like in the LCFSp queue, that as $\alpha \downarrow 1$ the average age approaches the lower bound $1/\lambda$. Similar observations are made for the log-normal distributed service~\eqref{eq:log_normal} and Weibull distributed service~\eqref{eq:Weibull}, we $\sigma \rightarrow +\infty$ and $\kappa \rightarrow 0$, respectively.

We now prove a simple lower bound on the average age, and show that the average age converges to this lower bound for the three heavy tailed service time distribution.
\begin{framed}
\begin{theorem}
\label{thm:opt2_gginf}
For the infinite server G/G/$\infty$ queue, the average age is lower-bounded by
\begin{equation}\nonumber
A^{\text{ave}}_{\text{G/G/}\infty}(F_S) \geq \frac{1}{2}\frac{\EX{X^2}}{\EX{X}}.
\end{equation}
Further, the lower-bound is achieved for
\begin{enumerate}
  \item Pareto distributed service~\eqref{eq:Par_Dist} as $\alpha \rightarrow 1$,
  \item Log-normal distributed service~\eqref{eq:log_normal} as $\sigma \rightarrow +\infty$, and
  \item Weibull distributed service~\eqref{eq:Weibull} as $\kappa \rightarrow 0$.
\end{enumerate}
\end{theorem}
\end{framed}
\begin{IEEEproof}
The lower-bound immediately follows from the average age expression in Lemma~\ref{lem:gginf}. We use a similar approach to that followed in the proof of Theorem~\ref{thm:LCFS_GG1_heavy_tail_opt}. We show that the same sufficient conditions as in Lemma~\ref{lem:suff_cond_lcfs} suffices for the average age optimality for the G/G/$\infty$ queue. %Since the heavy tailed service time distributions imply these conditions (see Lemma~\ref{}), we have the result.
\begin{framed}
\begin{lemma}
\label{lem:suff_cond_inf_serv}
Let a parametric, continuous, service time distribution, with parameter $\eta$, satisfy
\begin{enumerate}
  \item $\EX{S} = 1/\mu$,
  \item $\EX{\mathbb{I}_{\{ S > x\}}} \rightarrow 0$ as $\eta \rightarrow \eta^{\ast}$, and
  \item $\EX{S\mathbb{I}_{\{ S \leq x\}}} \rightarrow 0$ as $\eta \rightarrow \eta^{\ast}$,
\end{enumerate}
for some $\eta^{\ast}$, and all $x > 0$. Then
\begin{equation}
\lim_{\eta \rightarrow \eta^{\ast}} A^{\text{ave}}_{\text{G/G/}\infty}(F_S) = \frac{1}{2}\frac{\EX{X^2}}{\EX{X}}.
\end{equation}
\end{lemma}
\end{framed}
\begin{IEEEproof}
It suffices to argue that when the above conditions hold for a parametric random variable $S$, we have
\begin{equation}
\lim_{\eta \rightarrow \eta^{\ast}} \EX{\min_{l \geq 0}\left\{ \sum_{k=1}^{l}X_{k} + S_{l+1}\right\} } = 0.
\end{equation}
This is proved in Lemma~\ref{lem:gginf_iff_C} in Appendix~\ref{app:serv_dist_prop}.
\end{IEEEproof}

It, now, suffices to argue that the three heavy tailed service time distributions satisfy the conditions in Lemma~\ref{lem:suff_cond_inf_serv}. All the three heavy tailed distributions are continuous, and have mean $\EX{S} = 1/\mu$, by definition. The other two properties are verified in Appendix~\ref{pf:heavy_tail}.
\end{IEEEproof}

Thus, the minimum age can be achieved by the three heavy tailed service time distribution. For these three distributions, the second moment approaches infinity, as their tails get heavier; namely as $\alpha \rightarrow 1$, $\sigma \rightarrow +\infty$, and $\kappa \rightarrow 0$. This implies that the delay variance, which is lower-bounded by the variance in service time, also approaches infinity. However, it is not known that there is no other distribution that can simultaneously minimize average age and delay variance. In the next sub-section, we prove just that, and show that there is a strong age-delay variance tradeoff.

\subsection{Age-Delay Tradeoff}
\label{sec:inf_age_delay_tradeoff}

We now prove that there is a strong age-delay variance tradeoff.
\begin{framed}
\begin{theorem}
For the infinite server system, under the work conserving scheduling policy,
there is strong age-delay variance tradeoff.
\end{theorem}
\end{framed}
\begin{IEEEproof}
Let $A_{\min} = \frac{1}{2}\frac{\EX{X^2}}{\EX{X}}$ denote the minimum average age. We have to show that as $\AoI \rightarrow A_{\min}$ in~\eqref{eq:Inf_Aave_DelayVar_Tradeoff}, $V(\AoI) \rightarrow +\infty$. Note that the delay variance is lower-bounded by the variance in service time, namely, $\text{VarD}(F_S) \geq \EX{S^2} - \EX{S}^2$.
Therefore, it suffices to show that as $\AoI \rightarrow A_{\min}$ in~\eqref{eq:Inf_Aave_DelayVar_Tradeoff} we have $\EX{S^2} \rightarrow +\infty$.

To establish this, we use the average age expression derived in Lemma~\ref{lem:gginf}. Let $S_{\AoI}$ denote the service time, and $F_{S_{\AoI}}$ the corresponding service time distribution, that solves~\eqref{eq:Inf_Aave_DelayVar_Tradeoff}. Now, as $\AoI \rightarrow A_{\min}$ in~\eqref{eq:Inf_Aave_DelayVar_Tradeoff} we must have
\begin{multline}
A^{\text{ave}}_{\text{G/G/}\infty}(F_{S_{\AoI}}) = \frac{1}{2}\frac{\EX{X^2}}{\EX{X}} + \EX{\min_{l \geq 0}\left\{ \sum_{k=1}^{l}X_{k} + S_{l+1}\right\} } \\
\rightarrow \frac{1}{2}\frac{\EX{X^2}}{\EX{X}} = A_{\min},
\end{multline}
where $S_l$ and $X_k$ are independent and distributed according to $F_{S_{\AoI}}$ and $F_X$, respectively. This implies
\begin{equation}
\label{eq:xnuance1}
\lim_{\AoI \rightarrow A_{\min}} \EX{\min_{l \geq 0}\left\{ \sum_{k=1}^{l}X_{k} + S_{l+1}\right\} } = 0.
\end{equation}
From Lemma~\ref{lem:gginf_iff_C} and Lemma~\ref{lem:C_implies_S2}, in Appendix~\ref{app:serv_dist_prop},~\eqref{eq:xnuance1} implies that
\begin{equation}
\lim_{\AoI \rightarrow A_{\min}} \EX{S^{2}_{\AoI}} = +\infty,
\end{equation}
which proves the result.
\end{IEEEproof}

In the next sub-section, we prove an even stronger disparity between average age and delay. We show that the service time distribution that minimizes delay variance, i.e. deterministic service, yields the worst case age.

\subsection{Age-Delay Disparity}
\label{sec:inf_age_delay_disparity}

We first prove that deterministic service yields the worst average age, across all service time distributions.
\begin{framed}
\begin{theorem}
\label{thm:opt_gginf}
For the infinite server G/G/$\infty$ queue,
\begin{equation}\nonumber
A^{\text{ave}}_{\text{G/G/}\infty}(\lambda, \mu) \leq A^{\text{ave}}_{\text{G/D/}\infty}(\lambda, \mu),
\end{equation}
for all packet generation and service rates, $\lambda$ and $\mu$, respectively.
\end{theorem}
\end{framed}
\begin{IEEEproof}
See Appendix~\ref{pf:thm:opt_gginf}.
\end{IEEEproof}
The intuition is as follows: In the G/G/$\infty$ queue, packets do not get serviced in the same order as they are generated. However, a swap in order helps improve age, because it means that a packet that arrived later was served earlier. Therefore, the service that swaps the packet order the least maximizes age. Under deterministic service, the packet order is retained exactly, with probability $1$, and therefore, yields the maximizes age.

Notice that, for the G/G/$\infty$ queue, packet delay equals the service time, and therefore, deterministic service minimizes delay variance.
This observation, along with Theorem~\ref{thm:opt_gginf}, imply that for the G/G/$\infty$ queue, the \emph{service time distribution that reduces packet delay variance, maximizes average age of information}. %Furthermore, the heavy tailed service time distributions, that minimize average age, results in the worst case, unbounded, variance in packet delay; as $\EX{S^2} \rightarrow +\infty$.

The next section considers the general $M$ server system, and proves the age-delay tradeoff result of Theorem~\ref{thm:M_tradeoff}.

\section{$M$ Server System}
\label{sec:M_ServSyst}

In this section, we consider the $M$ server system and prove Theorem~\ref{thm:M_tradeoff}, which asserts a strong tradeoff between age-delay and age-delay variance. Recall that $A^{\text{ave}}(F_S, \pi_Q)$, $D(F_S, \pi_Q)$, and $\text{VarD}(F_S, \pi_Q)$ denote the average age, delay, and delay variance, respectively, under the scheduling policy $\pi_Q$ and the service time distribution $F_S$.

We first derive the minimum average age $A_{\min}$, over the space of all scheduling policies $\pi_Q$ and service time distributions $F_S$.
\begin{framed}
\begin{lemma}
\label{lem:AoI_min}
The minimum average age $A_{\min} = \frac{1}{2}\frac{\EX{X^2}}{\EX{X}}$.
\end{lemma}
\end{framed}
\begin{IEEEproof}
The fact that $\frac{1}{2}\frac{\EX{X^2}}{\EX{X}}$ is a lower-bound on the average age, can be proved by pretending that each update packet spends zero time in the system, i.e. $t_i = t^{'}_i$. This provides a sample path lower bound for the age process. In this sample path, the age drops to $0$ at every $t_i$, and increases to $t_{i+1} - t_{i}$, just before dropping to $0$ again at $t_{i+1}$. The average age of this artificially constructed age process is $\frac{1}{2}\frac{\EX{X^2}}{\EX{X}}$, and since it is a sample path wise lower-bound, we have $A^{\text{ave}}(F_S, \pi_Q) \geq \frac{1}{2}\frac{\EX{X^2}}{\EX{X}}$. This lower-bound $\frac{1}{2}\frac{\EX{X^2}}{\EX{X}}$ is independent of the scheduling policy $\pi_Q$ and the service time distribution $F_S$. Therefore, we have $A_{\min} \geq \frac{1}{2}\frac{\EX{X^2}}{\EX{X}}$.

In Section~\ref{sec:lcfs}, we showed that this lower-bound can be achieved by a single server system, i.e. $M = 1$, under the LCFSp scheduling policy with heavy tailed service. Therefore, choosing to route update packets only through a single sever, scheduling packets in that server with LCFSp scheduling policy with heavy tailed service, we can achieve this lower bound average age. Thus, $A_{\min} = \frac{1}{2}\frac{\EX{X^2}}{\EX{X}}$.
\end{IEEEproof}

%\begin{figure}
%  \centering
%  \includegraphics[width=\linewidth]{AoI_Delay_tradeoff2_plot2}
%  \caption{Plot of packet delay as a function of average age for LCFSp queue. Plots are for two service time distributions, namely, log-normal and Weibull. Packet generation is Poisson at rate $\lambda = 0.5$, and service rate $\mu = 0.8$.}
%  \label{fig:AoI_Delay_tradeoff2_plot2}
%\end{figure}
%In Figure~\ref{fig:AoI_Delay_tradeoff2_plot2}, we plot delay as a function of average age for the LCFSp queue scheduling discipline, and two service time distributions, namely log-normal and Weibull. Plotted curves are for different values of the distribution parameters $\sigma$ and $k$; see Table~\ref{tbl:heavy_tail}. The arrivals are Poisson at rate $\lambda = 0.5$, and therefore $A_{\min} = 1/\lambda = 2$.
%%
%We observe that in both the cases as $\sigma \rightarrow +\infty$ and $k \rightarrow 0$, respectively, the average age approaches $A_{\min}$, while delay goes to $+\infty$. This implies a strong age-delay tradeoff for the LCFSp queue scheduling discipline.
%%
%We now prove the strong age-delay and age-delay variance tradeoff, when the scheduling policy is not restricted to LCFSp.
%
We now prove the strong age-delay tradeoff and age-delay variance tradeoff.
\begin{framed}
\begin{lemma}
\label{thm:strong_age_delay}
For the $M$ server system, the following statements are true:
\begin{enumerate}
  \item When the update generation is Poisson, there is a strong age-delay tradeoff.
  \item When the update generation is a general renewal process, there is a strong age-delay variance tradeoff.
\end{enumerate}
\end{lemma}
\end{framed}
\begin{IEEEproof}
Let $A_{\min} = \frac{1}{2}\frac{\EX{X^2}}{\EX{X}}$ denote the minimum average age. Note that the variance in packet delay is lower-bounded by the variance in service time, under any scheduling policy $\pi_Q$. Therefore,
\begin{equation}
\label{eq:new_lb1}
\text{VarD}(F_S, \pi_Q) \geq \EX{S^2} - \EX{S}^2.
\end{equation}

It is known that the minimum delay can be attained by any work conserving scheduling policy~\cite{2016_ISIT_YinSun_AoI_Thput_Delay_LCFS, 2016_ISIT_TIT_YinSun_Thput_Delay_LCFS}. We define the following work conserving scheduling policy $\pi_{Q}^{\ast}$:
\begin{enumerate}
  \item Generated updates are queued in a single FCFS queue.
  \item Whenever a server is free, an update packet at the head of the FCFS queue, is assigned to that server.
\end{enumerate}
This scheduling policy, begin work conserving, attains minimum average delay for a given service time distribution, i.e.
\begin{equation}\label{eq:new_lb2_pre}
D(F_S, \pi_Q) \geq D(F_S, \pi_Q^{\ast}),
\end{equation}
for all scheduling policies $\pi_Q$. The $M$ server system under Poisson update generation and scheduling policy $\pi_Q^{\ast}$ is nothing but the M/G/k queue. For the M/G/k queue, the average delay, namely $D(F_S, \pi_Q^{\ast})$, is lower-bounded by a constant times the variance in service time~\cite{2010QS_gupta_MGk_Queue}:
\begin{equation}\label{eq:new_lb2}
D(F_S, \pi_Q^{\ast}) \geq c \left( \EX{S^2} - \EX{S}^2\right),
\end{equation}
where in~\cite{2010QS_gupta_MGk_Queue} the constant relates to the delay in M/M/k queue. From~\eqref{eq:new_lb2} and~\eqref{eq:new_lb2_pre}, we have
\begin{equation}\label{eq:new_lb3}
D(F_S, \pi_Q) \geq c \left( \EX{S^2} - \EX{S}^2\right),
\end{equation}
for any scheduling policy $\pi_Q$.

From~\eqref{eq:new_lb1} and~\eqref{eq:new_lb3}, it is clear, that in order to prove a strong age-delay tradeoff, for Poisson update generation, or to prove the strong age-delay variance tradeoff, for general update generation, it suffices to argue that $\EX{S^2} \rightarrow +\infty$. In the rest of the proof, we prove just this.

\textbf{1. Age-delay tradeoff:}
We consider Poisson update generation. Let $S_{\AoI}$ and $F_{S_{\AoI}}$ denote the service time random variable and its distribution, respectively, that solves~\eqref{eq:Aave_Delay_Tradeoff}. We argue that as $\AoI \rightarrow A_{\min}$ in~\eqref{eq:Aave_Delay_Tradeoff} we must have $\EX{S_{\AoI}^2} \rightarrow +\infty$.

We first note that the average age $A^{\text{ave}}(F_S, \pi_Q)$, under any queue scheduling policy $\pi_Q$, is lower-bounded by the average age for the G/G/$\infty$ queue:
\begin{equation}\label{eq:lb}
A^{\text{ave}}(F_S, \pi_Q) \geq A^{\text{ave}}_{\text{G/G/}\infty}(F_S).
\end{equation}
This is because, in G/G/$\infty$ queue, an arriving packet is immediately put to service, and therefore, incurs no queueing delay. Due to this the average age for the G/G/$\infty$ queue serves as a lower-bound for any $M$ server queue, in a stochastic ordering sense. Taking expected value yields~\eqref{eq:lb}.

We know the average age for the G/G/$\infty$ queue to be:
\begin{equation}\label{eq:Aave_gginf}
A^{\text{ave}}_{\text{G/G/}\infty} = \frac{1}{2}\frac{\EX{X^2}}{\EX{X}} + \EX{\min_{l \geq 0} \left( \sum_{k=1}^{l} X_k + S_{l+1} \right)},
\end{equation}
where $S_l$ and $X_k$ are independent random variables with distributions $F_{S_{\AoI}}$ and $F_X$, respectively.
Notice that the first term in~\eqref{eq:Aave_gginf} is nothing but $A_{\min}$. Therefore, as $\AoI \rightarrow A_{\min}$ in~\eqref{eq:Aave_Delay_Tradeoff}, it must be the case that $\EX{\min_{l \geq 0} \left( \sum_{k=1}^{l} X_k + S_{l+1} \right)} \rightarrow 0$. Lemmas~\ref{lem:gginf_iff_C} and~\ref{lem:C_implies_S2}, in Appendix~\ref{app:serv_dist_prop}, prove that $\EX{\min_{l \geq 0} \left( \sum_{k=1}^{l} X_k + S_{l+1} \right)} \rightarrow 0$ implies $\EX{S_{\AoI}^2} \rightarrow +\infty$.

\textbf{2. Age-delay variance tradeoff:} Let $S_{\AoI}$ and $F_{S_{\AoI}}$ denote the service time random variable and its distribution that solves~\eqref{eq:Aave_DelayVar_Tradeoff}. We have to argue that as $\AoI \rightarrow A_{\min}$ in~\eqref{eq:Aave_DelayVar_Tradeoff} we have $\EX{S_{\AoI}^2} \rightarrow +\infty$. We just proved this in establishing the age-delay tradeoff.
\end{IEEEproof}

%In the proof, we essentially showed that $\EX{S^2} \rightarrow +\infty$ is a necessary condition for the average age to approach the minimum $A_{\min}$. Results in Lemma~\ref{lem:AoI_min} and Lemma~\ref{thm:strong_age_delay} prove Theorem~\ref{thm:M_tradeoff}.
%

%\section{Discussion and Open Problems}
%\label{sec:discussion}

%\rt{State how the age-delay problem is a simplified one version of the age-delay tradeoff for a general network. The intuition presented in Section~\ref{} should hold also for a general network. Proving age-delay tradeoff for a general network, is therefore an open problem.}

\section{Conclusion}
\label{sec:conclusion}
We considered an $M$ server system in which each server serves at most one update packet at any given time. Updates are generated according to a renewal process and the system designer controls the scheduling discipline, routing, and the service time distribution. %For a fixed update generation and service rate we pose age-delay and age-delay variance tradeoff problems, which minimize packet delay and delay variance, respectively, subject to an average age constraint.
When the updates are generated according to a Poisson process, we show that there is a strong age-delay tradeoff, i.e. as the average age approaches its minimum the packet delay tends to infinity. However, for a general update generation process, we prove a strong age-delay variance tradeoff. The proof involves first establishing similar age-delay tradeoff results for two special cases of the $M$ server system, namely, the single server system with LCFSp service and the infinite server system. For the two cases, we also show that heavy tailed service time distributions asymptotically minimize average age, as their tail gets heavier.

Though seemingly counterintuitive, the age-delay tradeoff is natural and occurs due to the delays incurred by the non-informative packets. When the system attempts to minimize age, it minimizes waiting and service times for the informative packets. This results in very high waiting and service times for the non-informative packets, which dominate the packet delay, and causes it to increase unboundedly. We therefore expect similar age-delay tradeoffs to exist in other communication systems as well, and investigating them is an open question for future research.

%\rt{Change.}
%We considered the problem of minimizing age metrics over the space of packet generation and service time distributions. We showed that determinacy in update generation and service can yield the best or the worst case age, depending on the queueing system under consideration. While determinacy minimized age in the FCFS queue, for the LCFSp and G/G/$\infty$ queues, this was not necessarily the case.

%
%For the G/G/1 LCFSp queue and the infinite server G/G/$\infty$ queue, we showed that three heavy tailed service distributions, namely Pareto, log-normal, and Weibull, minimizes AoI metrics.
%
%For the M/G/1 LCFSp queue, we further showed that deterministic service, which minimizes packet delay, results in the worst case peak and average AoI.
%
%For the G/G/$\infty$ queue, we showed that deterministic service, which minimizes variance in packet delay, yields the worst case average AoI.
%
%Our results exposed a fundamental difference between packet delay and age metrics by showing that minimizing one can result in the worst case behavior for the other. We explore this difference further in~\cite{talak19_AoI_age_delay}.

\appendix

\subsection{Proof of Lemma~\ref{lem:LCFS_gg1}}
\label{pf:lem:LCFS_gg1}
Let $A(t)$ denote the age at time $t$. Let $B_i$ denote the age at the generation of the $i$th update packet, i.e. $Z_i = \sum_{k=0}^{i-1}X_k$:
\begin{equation}
B_{i} = A( Z_i ).
\end{equation}
Then, we have the following recursion for $B_i$:
\begin{equation}
B_{i+1} = \left\{ \begin{array}{cc}
                    X_i & \text{if}~S_i < X_i \\
                    B_i + X_i & \text{if}~S_i \geq X_i
                  \end{array}\right.,
\end{equation}
for all $i \geq 0$. This can be written as
\begin{equation}
B_{i+1} = X_i + B_{i}\left( 1 - \mathbb{I}_{S_i < X_i}\right).
\end{equation}
Note that $B_i$ is independent of $S_i$ and $X_i$. Further, $\{ B_i \}_{i \geq 1}$ is a Markov process, and can be shown to be positive recurrent using the drift criteria~\cite{meyn_markov_chains_stability}; using the fact that $X_i$ and $S_i$ are continuous random variables and $\pr{S_i < X_i} < 1$. Taking expected value, and noting that at stationarity $\EX{B_i} = \EX{B_{i+1}}$, we get
\begin{equation}\label{eq:B_exp}
\EX{B} = \frac{\EX{X}}{\pr{S < X}}.
\end{equation}

We now compute the average age. Let $R_i$ denote the area under the age curve $A(t)$ between the generation of packet $i$ and packet $i+1$:
\begin{equation}
R_i \triangleq \int_{Z_i}^{Z_i + X_i} A(t) dt,
\end{equation}
where $Z_i = \sum_{k=0}^{i-1}X_k$ is the time of generation of the $i$th update packet. This $R_i$ can be computed explicitly to be
\begin{equation}
R_i = \left\{ \begin{array}{cc}
                B_i X_i + \frac{1}{2}X_{i}^{2} & \text{if}~X_i < S_i \\
                B_i S_i + \frac{1}{2}X_{i}^{2} & \text{if}~X_i \geq S_i
              \end{array}\right.,
\end{equation}
which can be written compactly as
\begin{equation}\label{eq:Ri}
R_i = \frac{1}{2}X^{2}_{i} + B_{i}\min\left( X_i, S_i\right).
\end{equation}
Since, $B_i$ is independent of $X_i$ and $S_i$, taking expected value at stationarity we obtain
\begin{equation}\label{eq:R_exp}
\EX{R} = \frac{1}{2}\EX{X^2} + \EX{B}\EX{\min\left( X, S\right)}.
\end{equation}

Using renewal theory, the average age can be obtained to be
\begin{align}
A^{\text{ave}}_{\text{G/G/1}} &= \frac{\EX{R}}{\EX{X}} , \\
&= \frac{1}{2}\frac{\EX{X^2}}{\EX{X}} + \frac{\EX{B}}{\EX{X}}\EX{\min\left( X, S\right)}.
\end{align}
Substituting~\eqref{eq:B_exp} we get the result.

\subsection{Properties of the Heavy Tailed Distributions}
\label{pf:heavy_tail}

\begin{framed}
\begin{lemma}
For any $x > 0$, we have $\pr{S > x} \rightarrow 0$ and $\EX{S \mathbb{I}_{\{ S \leq x \}}} \rightarrow 0$ for:
\begin{enumerate}
  \item Pareto distributed service $S$, as $\alpha \rightarrow 1$; see~\eqref{eq:Par_Dist}.
  \item Log-normal distributed service $S$, as $\sigma \rightarrow +\infty$; see~\eqref{eq:log_normal}.
  \item Weibull distributed service $S$, as $\kappa \rightarrow 0$; see~\eqref{eq:Weibull}.
\end{enumerate}
\end{lemma}
\end{framed}
\begin{IEEEproof}

\textbf{1. Pareto Service:} Choose a $x > 0$. Then there exists a $\overline{\alpha}_x > 1$ such that $\theta(\alpha) = \frac{1}{\mu}\frac{\alpha - 1}{\alpha} < x$ for all $\alpha <  \overline{\alpha}_{x}$. For such any $\alpha <  \overline{\alpha}_{x}$, we have $\pr{S > x} = \left( \frac{\theta(\alpha)}{x} \right)^{\alpha} \rightarrow 0$ as $\alpha \downarrow 1$, since $\theta(\alpha) \rightarrow 0$ as $\alpha \downarrow 1$.

For the second part, we first compute $\EX{S\mathbb{I}_{\{ S \leq x\}}}$ for $\alpha < \overline{\alpha}_{x}$:
\begin{align}
\EX{S \mathbb{I}_{S \leq x}} &= \int_{\frac{1}{\mu}\left(1 - \frac{1}{\alpha}\right)}^{x} s f_{S}(s) ds = \frac{\alpha}{\mu^{\alpha}}\int_{\frac{1}{\mu}\left(1 - \frac{1}{\alpha}\right)}^{x} \frac{\left( 1 - \frac{1}{\alpha}\right)^{\alpha}}{s^{\alpha}} ds. \nonumber
\end{align}
Substituting $y = \alpha s/(\alpha - 1)$, and solving the definite integral, we get
\begin{equation}
\EX{S \mathbb{I}_{S \leq x}} =  \frac{1}{\mu} - \frac{1}{\mu} \frac{(\alpha/\mu)^{\alpha - 1}}{(\alpha - 1)^{\alpha - 1}} x^{\alpha - 1}.
\end{equation}
From the above expression, it can be deduced that $\EX{S \mathbb{I}_{S \leq x}} \rightarrow 0$ as $\alpha \downarrow 1$.

\textbf{2. Log-normal Service:} Choose a $x > 0$. From~\eqref{eq:log_normal} notice that
\begin{equation}\nonumber
\pr{S > x} = \pr{N > \frac{\log(x \mu)}{\sigma} + \frac{\sigma}{2}} \rightarrow 0,
\end{equation}
as $\sigma \rightarrow +\infty$.

For the second part, using the relation~\eqref{eq:log_normal} between the log-normal service time and normal random variable $N$, we can compute the expectation $\EX{S \mathbb{I}_{\{ S \leq x\}}}$ to be
\begin{equation}\nonumber
\EX{S \mathbb{I}_{\{ S \leq x\}} } = \frac{1}{\mu} - \frac{1}{\mu}\Phi\left( - \frac{\log(x \mu)}{\sigma} + \frac{\sigma}{2}\right),
\end{equation}
where $\Phi(x) = \frac{1}{\sqrt{2\pi}}\int_{-\infty}^{x}e^{-t^2/2}dt$. Taking the limit $\sigma \rightarrow +\infty$ we get $\Phi\left( - \frac{\log(x \mu)}{\sigma} + \frac{\sigma}{2}\right) \rightarrow 1$, and therefore, $\EX{S \mathbb{I}_{\{ S \leq x\}} } \rightarrow 0$.

\textbf{3. Weibull Service:} Choose a $x > 0$. Using the distribution function~\eqref{eq:Weibull}, we can conclude $\pr{S > x} = e^{-(x\mu)^{\kappa}}e^{-\left[ \Gamma(1 + 1/\kappa)\right]^{\kappa}}$. Using Sterling's formula, $\left[ \Gamma(1 + 1/\kappa)\right]^{\kappa} \geq 1/\kappa$, and therefore $\left[ \Gamma(1 + 1/\kappa)\right]^{\kappa} \rightarrow +\infty$ as $\kappa \rightarrow 0$. Therefore, we have $\pr{S > x} \rightarrow 0$ as $\kappa \rightarrow 0$.

For the second part, we can explicitly derive the conditional expectation $\EX{S\mathbb{I}_{\{S \leq x\}}}$ using the distribution~\eqref{eq:Weibull}:
\begin{align}
\EX{S \mathbb{I}_{\{ S \leq x \}}} &= \int_{0}^{x} \frac{\kappa}{\beta} \left( \frac{t}{\beta}\right)^{\kappa - 1} e^{- (t/\beta)^{\kappa}} t dt, \nonumber \\
&= \frac{1}{\mu \Gamma(1 + 1/\kappa)} \int_{0}^{\left(x \mu \Gamma(1 + 1/\kappa) \right)^{\kappa}} y^{1/\kappa} e^{-y} dy, \label{eq:no_to}
\end{align}
which is obtained by substituting $\beta = \left[ \mu \Gamma(1 + 1/\kappa)\right]^{-1}$ and changing variables $y = (t/\beta)^{\kappa}$. Using lower-bounds given by Sterling approximation on Gamma function, we can deduce that~\eqref{eq:no_to}, and therefore $\EX{S \mathbb{I}_{\{ S \leq x \}}}$, approaches $0$ as $\kappa \rightarrow 0$.
\end{IEEEproof}

\subsection{Proof of Theorem~\ref{thm:opt_LCFS_mg1}}
\label{pf:thm:opt_LCFS_mg1}
We first show that the average age for Poisson update generation is given by
\begin{equation}\label{eq:a0}
A^{\text{ave}}_{\text{M/G/1}} = \frac{\EX{S}}{\pr{S < X}}.
\end{equation}
Let $A(t)$ be the age at time $t$, and $B_i$ be the age at the time of generation of the $i$th update packet $Z_i = \sum_{k=0}^{i-1}X_k$:
\begin{equation}
B_{i} = A(Z_i).
\end{equation}
Let $B$ denote the distribution of $B_i$ at stationarity.
By PASTA property and ergodicity of the age process $A(t)$ we have $A^{\text{ave}}_{\text{M/G/1}} = \EX{B}$, as update generation process is a Poisson process. Substituting the expression for $\EX{B}$ in~\eqref{eq:B_exp}, from Appendix~\ref{pf:lem:LCFS_gg1}, we obtain~\eqref{eq:a0}.

Now, substituting $S = \EX{S}$ almost surely we get the average age expression for the $M/D/1$ LCFSp queue to be
\begin{equation}\label{eq:a1}
A^{\text{ave}}_{\text{M/D/1}} = \frac{\EX{S}}{\pr{\EX{S} < X}} = \frac{\EX{S}}{e^{-\lambda \EX{S}}},
\end{equation}
where we have used the fact that the packet inter-generation time $X$ is exponentially distributed. We obtain $A^{\text{ave}}_{\text{M/G/1}} \leq A^{\text{ave}}_{\text{M/D/1}}$ by noting that
\begin{equation}
\pr{S < X} = \EX{e^{-\lambda S}} \geq e^{-\lambda \EX{S}},
\end{equation}
by Jensen's inequality.

\subsection{Proof of Theorem~\ref{thm:opt_gginf}}
\label{pf:thm:opt_gginf}
From Lemma~\ref{lem:gginf}, it is clear that the average age depend on service time through the term:
\begin{equation}
\EX{\min_{l \geq 0}\left\{ \sum_{k=1}^{l}X_{k} + S_{l+1}\right\} }.
\end{equation}
We show that this quantity is maximized when service times are deterministic, i.e. $S = \EX{S}$ almost surely.

First, notice that
\begin{equation}
\min_{l \geq 0}\left\{ \sum_{k=1}^{l}X_{k} + S_{l+1}\right\} = S_{1},
\end{equation}
if $S_{k}$ are all equal and deterministic. This is because $X_k \geq 0$ almost surely. Thus, the peak and average age for the G/D/$\infty$ queue is given by
\begin{equation}
\label{eq:m0}
A^{\text{p}}_{\text{G/D/}\infty} = \EX{X} + \EX{S},
\end{equation}
and
\begin{equation}
\label{eq:m1}
A^{\text{ave}}_{\text{G/D/}\infty} = \frac{1}{2}\frac{\EX{X^2}}{\EX{X}} + \EX{S}.
\end{equation}
Furthermore, we must have
\begin{equation}
\min_{l \geq 0}\left\{ \sum_{k=1}^{l}X_{k} + S_{l+1}\right\} \leq S_{1},
\end{equation}
since $S_{1}$ is the first term in the minimization. Therefore,
\begin{equation}
\EX{\min_{l \geq 0}\left\{ \sum_{k=1}^{l}X_{k} + S_{l+1}\right\} } \leq \EX{S_{1}} = \EX{S}.
\end{equation}
Applying this to the peak and average age expression from Lemma~\ref{lem:gginf}, we get
\begin{equation}\label{eq:n0}
A^{\text{p}}_{\text{G/G/}\infty} \leq \EX{X} + \EX{S},
\end{equation}
and
\begin{equation}\label{eq:n1}
A^{\text{ave}}_{\text{G/G/}\infty} \leq \frac{1}{2}\frac{\EX{X^2}}{\EX{X}} + \EX{S}.
\end{equation}
The result follows from~\eqref{eq:m0},~\eqref{eq:m1},~\eqref{eq:n0}, and~\eqref{eq:n1}.

\subsection{Properties of Service Time Random Variable $S$}
\label{app:serv_dist_prop}
Here, we derive several asymptotic properties of the service time distributions and their implications. These properties are used throughout the paper.

Let $S$ be a continuous random variable with distribution $F_S$, with parameter $\eta$, such that $\EX{S} = 1/\mu$ for all $\eta$. For notational convenience, we hide the dependence of $S$ and $F_S$ on $\eta$.
We are interested in the $S$ and $F_S$ as $\eta \rightarrow \eta^{\ast}$, for some specific $\eta^{\ast}$.
\begin{framed}
\begin{lemma}
\label{lem:exists_to_all}
If $\exists$ a $x_0 > 0$ such that $\pr{S > x_0} \rightarrow 0$ and $\EX{S\mathbb{I}_{\{ S \leq x_0 \}}} \rightarrow 0$ as $\eta \rightarrow \eta^{\ast}$ then
\begin{equation}\label{eq:S_asymp_prop}
\lim_{\eta \rightarrow \eta^{\ast}} \pr{S > x} = 0~\text{and}~\lim_{\eta \rightarrow \eta^{\ast}} \EX{S\mathbb{I}_{\{ S \leq x \}}} = 0,
\end{equation}
for all $x \geq x_0$.
\end{lemma}
\end{framed}
\begin{IEEEproof}
Let there be a $x_0 > 0$ such that
\begin{equation}\label{eq:S_asymp_prop1}
\lim_{\eta \rightarrow \eta^{\ast}} \pr{S > x_0} = 0~\text{and}~\lim_{\eta \rightarrow \eta^{\ast}} \EX{S\mathbb{I}_{\{ S < x_0 \}}} = 0.
\end{equation}
Take a $x > x_0$. Then, $\mathbb{I}_{\{ S > x \}} \leq \mathbb{I}_{\{ S > x_0 \}}$, and therefore, $\pr{S > x} \leq \pr{S > x_0}$. This and~\eqref{eq:S_asymp_prop1} implies
\begin{equation}\label{eq:S_asymp_prop2}
\lim_{\eta \rightarrow \eta^{\ast}} \pr{S > x} = 0.
\end{equation}
For a $x > x_0$, we can re-write $\EX{S\mathbb{I}_{\{ S \leq x \}}}$ as
\begin{align}
\EX{S\mathbb{I}_{\{ S \leq x \}}} &= \EX{S\mathbb{I}_{\{ S \leq x_0 \}}} + \EX{S\mathbb{I}_{\{ x_0 < S \leq x \}}},  \\
   &\leq \EX{S\mathbb{I}_{\{ S \leq x_0 \}}} + x \EX{\mathbb{I}_{\{ x_0 < S \leq x \}}}, \\
   &\leq \EX{S\mathbb{I}_{\{ S \leq x_0 \}}} + x \pr{S > x_0}. \label{eq:S_asymp_prop3}
\end{align}
Using~\eqref{eq:S_asymp_prop1}, which states that both the terms in~\eqref{eq:S_asymp_prop3} tend to $0$ as $\eta \rightarrow \eta^{\ast}$, we get
\begin{equation}\label{eq:S_asymp_prop4}
\lim_{\eta \rightarrow \eta^{\ast}} \EX{S\mathbb{I}_{\{ S \leq x \}}} = 0.
\end{equation}

Since~\eqref{eq:S_asymp_prop2} and~\eqref{eq:S_asymp_prop4} hold for any $x > x_0$, we have the result.
\end{IEEEproof}

In the infinite server case, the average age expression in Lemma~\ref{lem:gginf} has a term
\begin{equation}
\EX{\min_{l \geq 0} \left( \sum_{k=1}^{l} X_k + S_{l+1} \right)},
\end{equation}
where $S_l$ and $X_k$ are independent, distributed according to $F_S$ and $F_X$, respectively.
We would like to derive conditions on $S$ such that
\begin{equation}\nonumber
\EX{\min_{l \geq 0} \left( \sum_{k=1}^{l} X_k + S_{l+1} \right)} \rightarrow 0,
\end{equation}
as $\eta$ approaches certain $\eta^{\ast}$,
for a given distribution $F_X$. The following result, derives an equivalent condition that only requires verifying certain properties of $F_S$.
\begin{framed}
\begin{lemma}
\label{lem:gginf_iff_C}
For $S_l$ and $X_k$ that are i.i.d. distributed according to $F_S$ and $F_X$, respectively, we have
\begin{equation}\label{eq:min_term}
\lim_{\eta \rightarrow \eta^{\ast}} \EX{\min_{l \geq 0} \left( \sum_{k=1}^{l} X_k + S_{l+1} \right)} = 0,
\end{equation}
if and only if, for all $x > 0$, we have
\begin{equation}\label{eq:suff_cond}
\lim_{\eta \rightarrow \eta^{\ast}} \pr{S > x} = 0,~\text{and}~\lim_{\eta \rightarrow \eta^{\ast}} \EX{S \mathbb{I}_{\{ S \leq x\}}} = 0.
\end{equation}
\end{lemma}
\end{framed}
\begin{IEEEproof}
\noindent \textbf{(a)}~We first prove that~\eqref{eq:min_term} implies~\eqref{eq:suff_cond}.
Let $Z = \min_{l \geq 0} \left( \sum_{k=1}^{l} X_k + S_{l+1} \right)$. We first lower-bound $Z$ as follow:
\begin{equation}\nonumber
Z = \min\{ S_1, X_1 + S_2, X_1 + X_2 + S_3, \ldots \} = \min\{ S_1, X_1 + Z' \},
\end{equation}
%\begin{align}
%Z &= \min\{ S_1, X_1 + S_2, X_1 + X_2 + S_3, \ldots \}, \nonumber  \\
%  &= \min\{ S_1, X_1 + Z' \}, \nonumber
%\end{align}
where $Z' = \min\{ S_2, X_2 + S_3, X_2 + X_3 + S_4, \ldots \}$. Since $Z' \geq 0$, we must have $Z \geq \min\{S_1, X_1\}$. Without loss of generality, we can loose the subscripts and write $Z \geq \min\{S, X\}$, where $S \sim F_S$ and $X \sim F_X$.

If $\EX{Z} \rightarrow 0$ as $\eta \rightarrow \eta^{\ast}$ then clearly $\EX{\min\{S, X\}} \rightarrow 0$ as $\eta \rightarrow \eta^{\ast}$. Pick a $x_0 > 0$ such that $\pr{X \geq x_0} > 0$. Note that such an $x_0 > 0$ always exists since $\EX{X} = 1/\lambda > 0$. Now construct $\hat{X}$ such that:
\begin{equation}\nonumber
\hat{X} = \left\{ \begin{array}{cc}
                    0 &~\text{if}~X < x_0 \\
                    x_0 &~\text{if}~X \geq x_0
                  \end{array}\right..
\end{equation}
Clearly, $\hat{X} \leq X$, and thus, $\min\{S, \hat{X}\} \leq \min\{S, X\}$, which implies $\EX{\min\{S, \hat{X}\}} \rightarrow 0$. Since $\hat{X}$ takes only two values, namely $0$ and $x_0$, we have $\EX{\min\{S, \hat{X}\}} = \EX{\min\{S, x_0 \}}\pr{X \geq x_0}$. Further, $\pr{X \geq x_0}$ does not depend on $S$, and therefore, is independent of the parameter $\eta$. Therefore, $\EX{\min\{S, \hat{X}\}} \rightarrow 0$ implies
\begin{equation}\label{eq:nuclear2}
\lim_{\eta \rightarrow \eta^{\ast}} \EX{\min\{S, x_0\}} = 0.
\end{equation}
Now, notice that
\begin{equation}\label{eq:nuclear3}
\EX{\min\{S, x_0\}} = \EX{S\mathbb{I}_{ \{ S \leq x_0 \} }} + x_0 \EX{\mathbb{I}_{ \{ S > x_0\} }},
\end{equation}
Substituting~\eqref{eq:nuclear3} in~\eqref{eq:nuclear2} we get
\begin{equation}\label{eq:nuclear4}
\lim_{\eta \rightarrow \eta^{\ast}} \EX{S\mathbb{I}_{ \{ S \leq x_0 \} }} = 0~\text{and}~\lim_{\eta \rightarrow \eta^{\ast}} \EX{\mathbb{I}_{ \{ S > x_0\} }} = 0.
\end{equation}
Using Lemma~\ref{lem:exists_to_all}, ~\eqref{eq:nuclear4} implies
\begin{equation}\label{eq:nuclear5}
\lim_{\eta \rightarrow \eta^{\ast}} \EX{S\mathbb{I}_{ \{ S \leq x \} }} = 0~\text{and}~\lim_{\eta \rightarrow \eta^{\ast}} \EX{\mathbb{I}_{ \{ S > x\} }} = 0,
\end{equation}
for all $x \geq x_0$.

Now, we had chosen $x_0$ to be such that $\pr{X \geq x_0} > 0$. Since $\EX{X} = 1/\lambda > 0$, the choice of $x_0$ could be as small, and close to $0$, as possible. This and~\eqref{eq:nuclear5} yield the result in~\eqref{eq:suff_cond}.

\noindent \textbf{(b)}~We now prove that~\eqref{eq:suff_cond} implies~\eqref{eq:min_term}. First, note that~\eqref{eq:suff_cond} along with the bounded convergence theorem~\cite{Durrett} imply
\begin{equation}\label{eq:suo1}
\lim_{\eta \rightarrow \eta^{\ast}} \pr{S > X} = 0~\text{and}~\lim_{\eta \rightarrow \eta^{\ast}} \EX{S \mathbb{I}_{\{ S \leq X \}}} = 0.
\end{equation}
Using the same arguments we also have
\begin{equation}\label{eq:suo2}
\lim_{\eta \rightarrow \eta^{\ast}} \EX{X \mathbb{I}_{\{ S > X\}}} = 0.
\end{equation}
Secondly, note that
\begin{equation}\label{eq:suo3}
\EX{\min_{l \geq 0} \left( \sum_{k=1}^{l} X_k + S_{l+1} \right)} \leq \EX{\min\{ S_1, X_1 + S_2 \}}.
\end{equation}
It suffices to show that $\EX{\min\{ S_1, X_1 + S_2 \}} \rightarrow 0$ as $\eta \rightarrow \eta^{\ast}$. To see this, we write $\EX{\min\{ S_1, X_1 + S_2 \}}$ as:
\begin{align}\nonumber
&\EX{ \min\{S_1, X_1 + S_2 \} } \nonumber \\
&= \EX{  S_1 \mathbb{I}_{\{ S_1 \leq X_1 \}}  } + \EX{\left[ X_1 + \min\{ S_1 - X_1, S_2\}\right]\mathbb{I}_{\{ S_1 > X_1\}} },\nonumber \\
&\leq  \EX{  S_1 \mathbb{I}_{\{ S_1 \leq X_1 \}}  } + \EX{\left[ X_1 + S_2\right]\mathbb{I}_{\{ S_1 > X_1\}} }, \nonumber \\
&= \EX{  S_1 \mathbb{I}_{\{ S_1 \leq X_1 \}}  } + \EX{X_1\mathbb{I}_{\{ S_1 > X_1\}}} + \EX{S_2 \mathbb{I}_{\{ S_1 > X_1\}}}, \nonumber \\
&\rightarrow 0,~~\text{as}~~\eta \rightarrow \eta^{\ast}, \nonumber
\end{align}
where the last equation follows from~\eqref{eq:suo1} and~\eqref{eq:suo2}.
\end{IEEEproof}

We now give a sufficient condition on the service time distributions $F_S$, parameterized by $\eta$, to have its second moment approach infinity. This result is used in proving the strong age-delay and age-delay variance tradeoffs.
\begin{framed}
\begin{lemma}
\label{lem:C_implies_S2}
For the parameterized, service time random variable $S$, we have $\lim_{\eta \rightarrow \eta^{\ast}} \EX{S^2} = +\infty$ if
\begin{equation}\label{eq:suff_cond2}
\lim_{\eta \rightarrow \eta^{\ast}} \pr{S > x} = 0,~\text{and}~\lim_{\eta \rightarrow \eta^{\ast}} \EX{S \mathbb{I}_{\{ S \leq x\}}} = 0,
\end{equation}
for all $x \geq x_0$, and some $x_0 > 0$.
\end{lemma}
\end{framed}
\begin{IEEEproof}
Let the two conditions~\eqref{eq:suff_cond2} hold for $S$. First, note that $\EX{S^2} \geq \EX{S^2 \mathbb{I}_{\{ S > x\}}} \geq x \EX{S \mathbb{I}_{\{ S > x\}}}$, for all $x > 0$. We can write $\EX{S \mathbb{I}_{\{ S > x\}}}$ as $\EX{S} - \EX{S \mathbb{I}_{\{ S \leq x\}}} \rightarrow 1/\mu$ as $\eta \rightarrow \eta^{\ast}$ by~\eqref{eq:suff_cond2} and the fact that $\EX{S} = 1/\mu$. Therefore, we have $\liminf_{\eta \rightarrow \eta^{\ast}} \EX{S^2} \geq x/\mu$ for all $x \geq x_0$. This can only be true if $\lim_{\eta \rightarrow \eta^{\ast}} \EX{S^2} = +\infty$.
\end{IEEEproof}

%\bibliographystyle{ieeetr}
%\bibliography{../../../../PaperTrack/delay-bib,../../../../PaperTrack/books-bib,../../../../PaperTrack/prob-bib,../../../../PaperTrack/cvxalgo-bib,../../../../PaperTrack/aoi-bib,../../../../PaperTrack/uavnet-bib,../../../../PaperTrack/cps-bib,../../../../PaperTrack/neelesh-bib,../../../../PaperTrack/opt-scheduling-bib,../../../../PaperTrack/commun-future-bib}

\end{document}